\documentclass[11pt,a4paper]{article}
\usepackage[a4paper,margin=1in]{geometry}
\usepackage{amsmath}
\numberwithin{equation}{section}
\usepackage{amssymb}
\usepackage{graphicx}
\usepackage{authblk}
\usepackage{hyperref}
\usepackage{placeins}
\usepackage{float}
\usepackage{relsize}
\usepackage{mathtools, nccmath}
\usepackage{xcolor}
\usepackage{sectsty}
\usepackage{tabularx}
\usepackage{array}
\usepackage{booktabs}
\usepackage{amsthm}
\graphicspath{{./}}
\usepackage{pgfplots}
\pgfplotsset{compat=1.18}

\title{The Tension of Space as Dark Energy: Dynamics and Phenomenology}

\author[1,2]{Muhammad Ghulam Khuwajah Khan\thanks{\href{mailto:b24bs1234@iitj.ac.in}{b24bs1234@iitj.ac.in}}\thanks{\href{mailto:khanmuhammadghulam@rjcollege.edu.in}{khanmuhammadghulam@rjcollege.edu.in}}\thanks{\href{mailto:dr.muhammad.ghulam.khuwajah@gmail.com}{dr.muhammad.ghulam.khuwajah@gmail.com}}}

\affil[1]{\textit{Department of Physics, Ramniranjan Jhunjhunwala College, Mumbai 400\,086, Maharashtra, India} \vspace{0.8em}}

\affil[2]{\textit{School of Artificial Intelligence and Data Science, Indian Institute of Technology Jodhpur, Jodhpur 342\,037, Rajasthan, India}}

\date{}

\begin{document}

\maketitle

\begin{abstract}
 We develop a phenomenological model in which the observed late-time dark-energy sector is interpreted as an effective residual tension of space. Motivated by recent observational hints that the dark-energy equation of state may exhibit mild low-redshift evolution, we begin with an intrinsic membrane description of spacetime and show that a uniform space tension contributes to the gravitational field equations with precisely the tensor structure of vacuum energy. We then consider a Dirac--Born--Infeld completion, which naturally introduces a hidden Abelian gauge sector. A late-time hidden transition, \(U(1)_h \, \to \, \mathbb{Z}_n\), can reorganize hidden magnetic energy into a coarse-grained flux-tube reservoir. If this reservoir exchanges energy with the dynamical part of the space-tension sector, the effective dark-energy density becomes time dependent and acquires a nontrivial equation of state. At the background level, the model gives a simple proof-of-concept realization of running dark energy and admits a transient crossing of the phantom divide. A restricted numerical scan shows that the best representative solution corresponds to a string-like hidden flux reservoir, with an effective low-redshift evolution that projects approximately to $(w_0^{\rm mod},w_a^{\rm mod})\simeq(-0.915,-0.457)$ in the Chevallier--Polarski--Linder plane. This does not exactly reproduce the compressed observational benchmark, but it generates a phenomenologically relevant departure from \(\Lambda\)CDM and illustrates how a late hidden-sector defect population can induce mild running of the residual vacuum-like tension of space. The construction is therefore best viewed as a physically transparent proof of concept rather than a complete theory of dark energy.
\end{abstract}

\newpage

\section{\centering Introduction}

When Einstein wrote the field equations of general relativity, he and many of his contemporaries believed that the universe was static. A direct application of the original field equations, however, does not naturally allow a static homogeneous universe with nonzero matter density. In 1917 Einstein therefore introduced the cosmological term \(\Lambda g_{\mu\nu}\) \cite{einstein1917}, leading to
\begin{equation}
 R_{\mu\nu} - \frac{1}{2} \, R \, g_{\mu\nu} + \Lambda g_{\mu\nu} \; = \; \frac{8\pi G}{c^4} T_{\mu\nu} \, .
\end{equation}
For a homogeneous and isotropic closed universe filled with pressureless matter of mass density \(\rho\), this term permits a static solution when
\begin{equation}
 \Lambda \; = \;  \frac{1}{a^2} \; = \; \frac{4\pi G\rho}{c^2} \, ,
\end{equation}
where \(a\) is the curvature radius. After Hubble's observational discovery of cosmic expansion \cite{Hubble1929} and the theoretical developments of Friedmann \cite{friedmann1922} and Lema\^{i}tre \cite{Lemaitre1927}, Einstein abandoned the original static-universe motivation for the cosmological term, reportedly calling it his biggest blunder.

The cosmological term returned to central importance at the end of the twentieth century. Two supernova projects, the Supernova Cosmology Project \cite{Perlmutter1999} and the High-Z Supernova Search Team \cite{Riess1998}, used Type Ia supernovae as standardizable candles and found them to be dimmer than expected in a decelerating universe. The resulting interpretation was that the expansion of the universe is accelerating. Independent support for late-time acceleration comes from Cosmic Microwave Background (CMB) anisotropies measured by DASI \cite{Halverson2002}, WMAP \cite{Bennett2003}, and Planck \cite{Planck2016}, as well as from Baryon Acoustic Oscillation (BAO) measurements \cite{Eisenstein2005, Percival2010}. For a classic review, see Ref.~\cite{Peebles2003}.

The simplest explanation of the observed acceleration is the standard \(\Lambda\)CDM model, in which the cosmological constant acts as a rigid dark-energy component with equation of state \(w = - 1\). A common phenomenological extension is the \(w\)CDM model, in which the dark-energy equation of state is still taken to be constant, but is allowed to differ from \(-1\). In this case the dark-energy pressure and density satisfy,

\begin{equation}
 p_{\, \rm DE} \; =\; w\rho_{\, \rm DE} \, c^2 \, ,
\end{equation}
with \(w\) treated as an additional cosmological parameter. A further extension is the Chevallier--Polarski--Linder (CPL) parametrization \cite{ChevallierPolarski2001, Linder2003},

\begin{equation}
 w(a) \; = \; w_0 + w_a(1 - a) \, ,
\end{equation}
which allows a simple two-parameter description of low-redshift dark-energy evolution. These models provide useful observational benchmarks against which proposed dynamical dark-energy mechanisms can be compared.

Many theoretical ideas have been explored in order to understand the dark-energy sector. Representative proposals include quintessence \cite{ratra1988, caldwell1998}, k-essence \cite{armendariz2000, armendariz2001}, modified gravity \cite{Carroll:2003wy, DeFelice:2010, Clifton:2011jh, Hinterbichler:2011tt, Mukohyama:2020ohe}, the Dvali--Gabadadze--Porrati braneworld \cite{dvali2000, deffayet2002}, Chaplygin gas models \cite{kamenshchik2001}, holographic dark energy \cite{cohen1999, hsu2004, li2004}, quantum-cosmological approaches \cite{Hawking1984, Duff1989, Wu2008}, dynamical relaxation of the cosmological constant through membrane nucleation events of top forms \cite{BrownTeitelboim1987}, and multiple top-form scenarios \cite{BrownTeitelboim1988}. For broad reviews, see Refs.~\cite{copeland2006, li2011}.

The present work is motivated by the possibility that the late-time dark-energy sector may not be exactly constant. Recent observational developments have shifted part of the focus from the old question of why the vacuum energy is so small to the more phenomenological question of whether the residual late-time dark-energy component is perfectly rigid. The first-year cosmological results from the Dark Energy Spectroscopic Instrument (DESI), based on BAO measurements over a wide redshift range, were found to be compatible with the standard \(\Lambda\)CDM model, but also showed a mild preference for a dark-energy component that evolves with time \cite{DESI2024VI}. This preference became more pronounced when DESI measurements were interpreted jointly with CMB and Type Ia supernova data, and subsequent DESI full-shape analyses continued to favor an evolving dark-energy component \cite{DESI2024VII, Calderon2024}. More recent DESI analyses based on the larger three-year data set further indicate that mild low-redshift evolution of the dark-energy sector remains a viable and potentially important possibility \cite{DESI2025DR2, Lodha2025}.

If this observational trend persists, then late-time cosmic acceleration may not be sourced by a perfectly rigid cosmological constant, but by an effective vacuum-like sector that evolves slowly over cosmological time. This motivates the search for physically transparent phenomenological descriptions of dynamical dark energy. In the present work we explore one such possibility. Instead of introducing an \textit{ad hoc} dark-energy fluid from the outset, we interpret the observed residual vacuum sector as the intrinsic tension of space itself.

The scope of the present work is deliberately modest. We do not attempt to solve the full cosmological constant problem, nor do we explain from first principles why the absolute vacuum scale is so small. We assume that the large microscopic contributions to the vacuum energy have already been suppressed or neutralized by some ultraviolet mechanism. Possible examples include vacuum-energy sequestering, Hawking-type Euclidean arguments for the cosmological term, four-form flux discretuum scenarios in which many flux sectors allow an exceptionally small effective \(\Lambda\) to emerge, and self-sustained superfluid-like vacuum models in which the effective vacuum energy dynamically adjusts toward equilibrium; see, for example, Refs.~\cite{Hawking1984, Duff1989, Wu2008, KaloperPadilla2014a, KaloperPadilla2014b, KaloperPadilla2015, KaloperEtAl2016, KaloperPadilla2017, DAmico2017, BoussoPolchinski2000, Volovik2005CC, KlinkhamerVolovik2008DynamicVacuum, KlinkhamerVolovik2008SelfTuning, KlinkhamerVolovik2009GluonicVacuum, KlinkhamerVolovik2010TowardsSolution, Volovik2013SuperfluidUniverse}. Our aim is instead to investigate whether the small residual dark-energy sector that survives at late times can be described effectively as a running tension of space.

This viewpoint connects the vacuum-like contribution in Einstein's equations to an intrinsic geometric property of spacetime. A uniform tension of space behaves like a cosmological constant. However, if the intrinsic tension sector is supplemented by additional hidden-sector structure, then the total effective tension need not remain exactly constant. In particular, a late-time hidden transition can generate a defect-like reservoir that exchanges energy with the dynamical part of the space tension, thereby inducing a time-dependent dark-energy density and an effective equation of state different from \(-1\).

The roadmap of the paper is as follows. In Section~2 we formulate the intrinsic membrane picture in its minimal form. Starting from the standard Nambu--Goto action for extended objects, we specialize to the case in which the three-brane is identified with physical space itself, and show explicitly that a uniform space tension contributes to the field equations with exactly the tensor structure of vacuum energy. We then generalize this intrinsic description to its Dirac--Born--Infeld completion, which naturally introduces a hidden Abelian gauge sector on the membrane of space. In Section~3 we investigate a late-time effective origin for the running of the space tension. We show how a hidden \(U(1)_h \to \mathbb{Z}_n\) breaking can generate a coarse-grained flux-tube reservoir, how the total tension sector may be split into a rigid background piece and a dynamical contribution, and how energy exchange between the two leads to a time-dependent effective dark-energy equation of state. In Section~4 we present the observational benchmarks and results, comparing the resulting running-tension evolution with effective CPL constraints inferred from current late-time cosmological data. Finally, in Section~5 we discuss the physical interpretation of the construction, its limitations, and possible directions for future work. A complementary companion work, Ref.~\cite{KhanSpatialPhonons2025}, develops an alternative late-time realization of the same general picture of dynamical dark energy, in which the residual tension of space is supplemented not by a hidden defect reservoir but by a viscous longitudinal phonon fluid living on the elastic brane.

We now begin with the minimal intrinsic description in which space is treated as an elastic membrane with a Nambu--Goto type volume action.

\section{\centering Modeling Space as an Elastic Membrane}

\subsection{\centering Review of Nambu--Goto Dynamics for Strings and Branes}

A fundamental string in $1+1$ dimensions has worldsheet action \cite{Goto1971},

\begin{equation}
 S_1 = -T_1 \int d^2\xi \,\sqrt{-\det(\gamma_{\alpha\beta})}
\end{equation}
with worldsheet coordinates $\xi^\alpha=(\tau,\sigma)$, string tension $T_1$ and induced metric,

\begin{equation}
 \gamma_{\alpha\beta} = \frac{\partial X^\mu}{\partial \xi^\alpha} \frac{\partial X^\nu}{\partial \xi^\beta} g_{\mu\nu}
 \label{tensor transformation equation}
\end{equation}
For a $d$ brane in $(d{+}1)$ dimensions the natural generalization is \cite{Polchinski:1998rq, Achucarro:1987nc},

\begin{equation}
 S_d = - T_d \int d^{d+1}\xi \,\sqrt{(-1)^d \det(\gamma_{\alpha\beta})}
\end{equation}
where $T_d$ has dimensions of energy density in $d$ spatial dimensions. For $d \ge 3$ it is common to call $T_d$ a volume tension, since it generalizes surface tension to higher codimension.

We will specifically focus on the $d = 3$ case,

\begin{equation}
 S_3 = -T_3 \int d^4 \xi \, \sqrt{-\det(\gamma_{\alpha\beta})}
 \label{NG action of 3 brane}
\end{equation}

\subsection{\centering Stress Tensor of an Infinitely Extended $D_3$ Brane}

The stress tensor that couples to the ambient metric is defined by,

\begin{equation}
 T^{\mu\nu}_{(3)}(x) = \frac{2}{\sqrt{-g(x)}} \, \frac{\delta S_3}{\delta g_{\mu\nu}(x)}
\end{equation}
To vary $S_3$, note the standard determinant identity,

\begin{equation}
 \delta\sqrt{-\det\gamma} = \frac12 \sqrt{- \det\gamma} \, \gamma^{\alpha\beta} \delta\gamma_{\alpha\beta}
\end{equation} 
Using $\delta\gamma_{\alpha\beta} = \partial_\alpha X^\mu \partial_\beta X^\nu \,\delta g_{\mu\nu}$, one finds,

\begin{equation}
 \frac{\delta S_3}{\delta g_{\mu\nu}(x)} = - \frac{T_3}{2} \int d^4 \xi \, \sqrt{-\det\gamma} \, \gamma^{\alpha\beta} \, \frac{\partial X^\mu}{\partial \xi^\alpha} \frac{\partial X^\nu}{\partial \xi^\beta} \, \delta^{(4)} \big(x - X(\xi)\big)
\end{equation}
Thus,

\begin{equation}
 T^{\mu\nu}_{(3)}(x) = - \frac{T_3}{\sqrt{-g(x)}} \int d^4 \xi \, \sqrt{-\det\gamma} \, \gamma^{\alpha\beta} \, \frac{\partial X^\mu}{\partial \xi^\alpha} \frac{\partial X^\nu}{\partial \xi^\beta} \, \delta^{(4)} \big(x - X(\xi)\big)
 \label{stress energy tensor equation}
\end{equation}

For an infinitely extended brane the natural gauge choice is the static gauge. This gauge identifies the time coordinate $\tau$ and the spatial coordinates $\sigma^1,\sigma^2,\sigma^3$ on the brane's worldsheet with the coordinate time $t$ and spatial coordinates $x,y$ and $z$ of the target spacetime. Equivalently, the spacetime embedding coordinates of the brane become $X^0 = \tau$ and $X^i = \sigma^i$ for $i = 1, 2, 3$. Therefore, in the static gauge, we have,

\begin{equation}
 \frac{\partial X^\mu}{\partial \xi^\alpha} = \delta^\mu_\alpha
 \label{static gauge choice}
\end{equation}
where $\delta^\mu_\alpha$ is the Kronecker delta. Furthermore, equation \eqref{tensor transformation equation} dictates that the induced metric $\gamma_{\alpha \beta}$ on the brane world-volume is related to the target spacetime metric $g_{\mu\nu}$ as,

\begin{equation}
 \gamma_{\alpha \beta} = g_{\mu \nu} \frac{\partial X^\mu}{\partial \xi^\alpha} \frac{\partial X^\nu}{\partial \xi^\beta}
\end{equation}
Substituting \eqref{static gauge choice} in the equation above, we obtain,

\begin{equation}
 \gamma_{\alpha \beta} = g_{\mu \nu}\ \delta^\mu_\alpha \ \delta^\nu_\beta
\end{equation}
which reduces to, 

\begin{equation}
 \gamma_{\alpha \beta} = g_{\alpha \beta}
\end{equation}
Therefore, we see that the induced metric $\gamma_{\alpha\beta}$ on the $3$-brane world-volume is equal to the target spacetime metric $g_{\alpha \beta}$. Furthermore, the equality above directly implies that $\sqrt{-\det(\gamma_{\alpha\beta})}=\sqrt{-g}$ and $\gamma^{\alpha \beta}=g^{\alpha \beta}$. This simplifies the expression for the $3$-brane stress energy tensor  $T^{\mu\nu}_{(3)}$ (equation \eqref{stress energy tensor equation}) to,

\begin{equation}
 T^{\mu\nu}_{(3)} = - T_3 \int d^4\xi \ g^{\alpha\beta} \frac{\partial X^\mu}{\partial \xi^\alpha} \frac{\partial X^\nu}{\partial \xi^\beta} \delta^{(3+1)}(x - X^\mu(\xi))
\end{equation}
In order to evaluate the expression above, we can rewrite it as,

\begin{equation}
 T^{\mu\nu}_{(3)} = - T_3 \int_{-\infty}^{\infty} d\tau \int_{-\infty}^{\infty} d^3\sigma\ g^{\alpha\beta} \frac{\partial X^\mu}{\partial \xi^\alpha} \frac{\partial X^\nu}{\partial \xi^\beta} \delta(x^0 -\tau) \prod_1^3 \delta(x^i - \sigma^i)
\end{equation}
which allows us to straightforwardly perform the integral with respect to $\tau$ and obtain,

\begin{equation}
 T^{\mu\nu}_{(3)} = - T_3 \int_{-\infty}^{\infty} d^3\sigma\ g^{\alpha\beta} \frac{\partial X^\mu}{\partial \xi^\alpha} \frac{\partial X^\nu}{\partial \xi^\beta} \prod_1^3 \delta(x^i - \sigma^i)
\end{equation}
Furthermore, given that we have assumed that the brane is infinitely large, we have,

\[\displaystyle \int_{-\infty}^{\infty} d^3\sigma \prod_1^3 \delta(x^i - \sigma^i) = 1\] 
since $x^i$ is always within the range of $\sigma^i$ $\forall \ i$. Therefore we obtain,

\begin{equation}
 T^{\mu\nu}_{(3)} = - T_3 \ g^{\alpha\beta} \frac{\partial X^\mu}{\partial \xi^\alpha} \frac{\partial X^\nu}{\partial \xi^\beta}
\end{equation}
Using equation \eqref{static gauge choice} in the equation above, we obtain,

\begin{equation}
 T^{\mu\nu}_{(3)} = - T_3 \ g^{\alpha \beta}\ \delta^\mu_\alpha \ \delta^\nu_\beta
\end{equation}
which gives

\begin{equation}
 T^{\mu\nu}_{(3)} = - T_3 \ g^{\mu \nu}
\end{equation}
and finally, we obtain,

\begin{equation}
 T_{\mu\nu(3)} = - T_3 \ g_{\mu\nu}
\end{equation}
Comparing with the vacuum stress tensor,

\begin{equation}
 T_{\mu\nu}^{\rm vac} \; = \; - \rho_{\rm vac}c^2 g_{\mu\nu},
\end{equation}
we see that a uniform three-brane tension \(T_3\) behaves gravitationally like a vacuum energy density, with the identification \(T_3 = \rho_{\rm vac}c^2\).

\subsection{\centering Space as a $D_3$ Brane}

The previous derivation invites a sharper hypothesis. Instead of positing a separate embedded brane, we may treat \emph{space itself} as a three-dimensional elastic medium with an intrinsic volume tension $T_s$. The worldvolume of space is the physical spacetime and its worldvolume metric is simply $g_{\mu\nu}$. The Nambu--Goto-like action for space is then,

\begin{equation}
 S_s = -T_s \int d^4x \,\sqrt{-g}
 \label{action of space}
\end{equation}
This has the same dimensions as equation \eqref{NG action of 3 brane} with $T_3$ replaced by $T_s$. It also perfectly mirrors the cosmological piece of the Einstein--Hilbert action. The latter reads,

\begin{equation}
 S_{\mathrm{EH}} = \int d^4x \sqrt{-g}\left(\frac{c^4}{16\pi G}\,(R - 2\Lambda) + \mathcal L_M\right) 
\end{equation}
in which the cosmological piece is,

\begin{equation}
 S_{\Lambda} = -\frac{\Lambda c^4}{8\pi G}\int d^4x \sqrt{-g}
\end{equation}
Using $\Lambda = 8\pi G\,\rho_{\mathrm{vac}}/c^2$ in the equation above, we obtain, 

\begin{equation}
 S_{\Lambda} = -\rho_{\mathrm{vac}} c^2 \int d^4x \sqrt{-g}
\end{equation}
Hence, dimensionally $[\rho_{\mathrm{vac}} c^2] = [T_s] = [T_3]$, and $S_s$ given by equation \eqref{action of space} exactly mimics $S_{\Lambda}$ at the level of local dynamics.

We now examine the stress tensor that follows from the elastic action of space, that is equation \eqref{action of space}. On varying $S_s$, we obtain,

\begin{equation}
 \delta S_s = -T_s \int d^4x \, \delta\sqrt{-g} = -\frac{T_s}{2} \int d^4x \, \sqrt{-g} \, g^{\mu\nu}\delta g_{\mu\nu}
\end{equation}
Therefore,

\begin{equation}
 \frac{2}{\sqrt{-g}} \, \frac{\delta S_s}{\delta g_{\mu\nu}} \; = \; - T_s g^{\mu\nu}.
\end{equation}
But, the left hand side is the definition of the stress tensor for space, that is,

\begin{equation}
 T^{\mu\nu}_{(s)} = \frac{2}{\sqrt{-g}} \, \frac{\delta S_s}{\delta g_{\mu\nu}}
\end{equation}
Therefore, we have,

\begin{equation}
 T^{\mu\nu}_{(s)} = - T_s \, g^{\mu\nu}
\end{equation}
or that,
\begin{equation}
 T_{\mu\nu(s)} = - T_s \, g_{\mu\nu}
\end{equation}
Thus the elastic model of space reproduces the vacuum stress tensor with the identification $T_s = \rho_{\mathrm{vac}} \, c^2$. Equivalently, we have,

\begin{equation}
 T_s = \frac{\Lambda c^4}{8 \pi G} \, .
 \label{relation between T and Lambda}
\end{equation}
In the present work, we take the next step and identify this intrinsic tension of space directly with the observed late-time vacuum sector. More precisely, \(T_s\) is not interpreted as the naive bare zero-point energy of all ultraviolet degrees of freedom. Rather, it is understood as the small residual vacuum energy density that remains after the ultraviolet physics responsible for suppressing the enormous microscopic contributions has already done its job as mentioned earlier. Our standpoint here is deliberately agnostic about which ultraviolet mechanism is ultimately realized in nature. We simply take the observed small cosmological term as an infrared input and identify it with the effective volume tension of space itself. In this sense, \(T_s\) should be viewed as the residual vacuum sector that survives after the ultraviolet vacuum-energy problem has been dealt with, rather than as a first-principles derivation of that suppression within the present model.

Therefore, for $\Lambda \approx 1.09 \times 10^{-52} \, \mathrm{m}^{-2}$ \cite{Planck2018Cosmo}, equation \eqref{relation between T and Lambda} gives the tension of space as,

\begin{equation}
 T_s \approx 5.249 \times 10^{-10} \; \mathrm{J/m^{3}}
\end{equation}
Furthermore, we do not assume that physical space is embedded in a higher dimensional ambient manifold. The elastic membrane language in this paper is an intrinsic description and all observables are built from the intrinsic metric and its derived tensors. Introducing an embedding is possible as a change of variables but it adds redundant structure that is not needed for the dynamics we study. An embedding also introduces an arbitrary ambient metric that does not enter any measurement performed by observers confined to the physical spacetime. Our equations of motion are fully determined by intrinsic geometry, so the theory remains sensible without any reference to extrinsic data. For these reasons we avoid committing to an embedding and keep the framework intrinsic from the outset.

The construction developed so far captures only the uniform tension sector of the intrinsic membrane. However, once the membrane of space is allowed to support an internal hidden $U(1)$ gauge field, the Nambu--Goto description is no longer the most complete local nonlinear effective action. The natural generalization is then of Dirac--Born--Infeld type. In the limit that the hidden gauge field is switched off, the DBI action reduces to the Nambu--Goto volume term. When the gauge field is present, it resums the nonlinear response of the membrane to finite worldvolume fluxes and provides a finite-field completion of the Maxwell regime. We therefore turn next to the DBI completion of the intrinsic membrane picture and ask what kind of hidden magnetic sectors it admits.

\subsection{\centering The \textit{DBI} Completion of the Elastic Space Picture}

A convenient starting point is the four-dimensional DBI gauge action (we work in units where $\hbar = c = 1$),

\begin{equation}
 S_{\rm DBI} = - T_s \int d^4x \, \sqrt{-\det \left(g_{\mu\nu} +\lambda_h F_{\mu\nu}\right)} \, .
\end{equation}
Factoring out the metric determinant gives,

\begin{equation}
 S_{\rm DBI} = -T_s \int d^4x \, \sqrt{-g} \, \sqrt{\det \left(\delta^\mu{}_\nu + \lambda_h F^\mu{}_\nu\right)} \, .
\end{equation}
Since \(F_{\mu\nu}\) is antisymmetric, the determinant identity in four dimensions yields,

\begin{equation}
 \det \left(\delta^\mu{}_\nu + \lambda_h F^\mu{}_\nu\right) = 1 + \frac{\lambda_h^2}{2}F_{\mu\nu} \, F^{\mu\nu} - \frac{\lambda_h^4}{16} (F_{\mu\nu} \tilde F^{\mu\nu})^2 \, .
\end{equation}
Therefore,

\begin{equation}
 S_{\rm DBI} = - T_s \int d^4 x \, \sqrt{-g} \, \sqrt{1 + \frac{\lambda_h^2}{2} F_{\mu\nu} \, F^{\mu\nu} - \frac{\lambda_h^4}{16} (F_{\mu\nu} \tilde F^{\mu\nu})^2} \, .
\end{equation}
Writing \(\lambda_h = \beta_h^{-1}\), one obtains,

\begin{equation}
 S_{\rm DBI} = - T_s \int d^4x \, \sqrt{-g} \, \sqrt{1 + \frac{1}{2\beta_h^2} \, F_{\mu\nu} \, F^{\mu\nu} -\frac{1}{16\beta_h^4} \, (F_{\mu\nu} \tilde F^{\mu\nu})^2} \, .
\end{equation}
Note that the equation above in weak fields reduces to,

\begin{equation}
 S_{\rm DBI} = - T_s \int d^4x \, \sqrt{-g} \, \left[1 + \frac{1}{4 \, \beta_h^2} \, F_{\mu\nu} \, F^{\mu\nu} \right] \, .
\end{equation}
 When \(\beta_h^2 = T_s\) as in the Maxwell normalization, we have,

\begin{equation}
 S_{\rm DBI} = - \int d^4x \, \sqrt{-g} \, \left[T_s + \frac{1}{4} \, F_{\mu\nu} \, F^{\mu\nu} \right] \, ,
 \label{reduced action}
\end{equation}
with the Lagrangian density,

\begin{equation}
 \mathcal{L}_{\rm DBI} = - T_s - \frac{1}{4} F_{\mu\nu} \, F^{\mu\nu} \, .
 \label{Minimal Lag}
\end{equation}
At this point, the minimal physical content of the construction becomes clear. Once space is treated as an effective DBI membrane, the theory already contains two basic ingredients. The first is the vacuum-like or cosmological contribution \(T_s\), which in the present framework is identified with the residual observed vacuum sector. The second is an abelian gauge sector arising from the DBI structure itself, whose weak-field limit is the \(U(1)\) field appearing above. In this work, we regard that \(U(1)\) as hidden. The reason is simple. Nothing in the bare effective action requires it to be identified with ordinary electromagnetism, and in particular the visible matter content and coupling assignments do not follow automatically from the minimal DBI sector alone. One may certainly imagine more elaborate ultraviolet completions in which Standard Model gauge fields and matter are engineered from intersecting or stacked brane configurations, but such constructions are highly model dependent and often rather contrived. Our purpose here is more modest. We only use the minimal DBI completion as an intrinsic effective description of space, which naturally furnishes a vacuum term together with a hidden \(U(1)\) sector, without assuming that this sector is the electromagnetic field of the Standard Model.

\section{\centering Late-Time Running of the Tension of Space}

We now consider a late-time microscopic origin for the running of the
space tension. The purpose of the present section is not to provide a complete ultraviolet model of hidden-sector interaction. Rather, we construct a proof of concept effective description in which the intrinsic hidden gauge sector already suggested by the DBI completion of space develops, \textit{at late times}, a defect-like reservoir that can exchange energy with the vacuum-like tension sector. We consider here an additional hidden-sector transition at late times, whose detailed microphysics we do not attempt to derive here, but whose coarse-grained cosmological effects can still be studied consistently.

\subsection{\centering Hidden $U(1)\to \mathbb{Z}_n$ breaking and Flux-Tube Defects}

We now extend the intrinsic hidden $U(1)$ gauge sector of space, already
introduced in the DBI description, by a complex scalar field $\Phi$ carrying
charge $n$ under that same hidden Abelian symmetry. At the level of a minimal effective description, the hidden sector is modeled by extending the DBI Lagrangian of space, equation \eqref{Minimal Lag}, so as to include a hidden Higgs field. The corresponding effective Lagrangian is

\begin{equation}
 \mathcal{L}_{\rm hid} = -\frac{1}{4} Z(\sigma) \, F_{\mu\nu} \, F^{\mu\nu} + |D_\mu \Phi|^2 - \frac{\lambda_\Phi}{4} \left( |\Phi|^2 - v_\Phi^2(\sigma) \right)^2 \, , 
 \label{eq:hidden_U1_Zn_L}
\end{equation}
with,

\begin{equation}
 D_\mu \Phi = \left( \partial_\mu - i \, n e_h A_\mu \right)\Phi \, .
 \label{eq:hidden_cov_deriv}
\end{equation}
Here $A_\mu$ is the hidden gauge field, $e_h$ is the hidden gauge coupling, and $\sigma$ is a slow collective variable that will be used below to parameterize the dynamical part of the space-tension sector. The functions $Z(\sigma)$ and $v_\Phi(\sigma)$ allow the hidden sector to communicate with the tension sector at the effective level.

When the hidden Higgs field condenses, at a late enough time in cosmic history (as discussed in Sec.~4.2),

\begin{equation}
 \langle \Phi \rangle = \frac{v_\Phi}{\sqrt{2}} \neq 0 \, ,
\end{equation}
the continuous hidden gauge symmetry is not necessarily destroyed completely. Because $\Phi$ carries charge $n$, the vacuum remains invariant under the discrete subgroup $\mathbb{Z}_n \subset U(1)_h$ \cite{KraussWilczek1989, PreskillKrauss1990}. At the effective level, the late-time symmetry-breaking pattern is therefore,

\begin{equation}
 U(1)_h \longrightarrow \mathbb{Z}_n \, .
 \label{eq:U1_to_Zn}
\end{equation}
Owing to this condensation, the late-time hidden sector enters a topologically nontrivial broken phase in which localized flux-tube defects can form, see \cite{NielsenOlesen1973, Kibble1976, HindmarshKibble1995, VilenkinShellard1994}. The basic mechanism is standard. During the transition, different causal regions need not select the same phase of the hidden complex scalar $\Phi$. As a result, the phase of $\Phi$ can acquire a nontrivial winding around certain closed loops in space. Such a winding cannot be removed continuously while remaining everywhere in the broken phase. At least along some line-like locus, the scalar magnitude must therefore be driven back toward zero so that the phase becomes ill-defined there. These line-like loci are precisely the cores of the hidden flux tubes.

The gauge field then adjusts so as to maintain finite energy away from the core. In particular, the covariant gradient energy of the Higgs field requires the phase winding of $\Phi$ to be compensated by an azimuthal gauge configuration, so that the hidden magnetic flux becomes trapped inside the defect core rather than remaining as a freely propagating long-range field. Equivalently, once the hidden $U(1)$ sector is Higgsed to a residual discrete subgroup, the gauge boson acquires a mass in the broken phase, and the bulk expels magnetic flux in the usual Meissner-like manner \cite{ NielsenOlesen1973, Abrikosov1957, Tinkham2004}. The flux is therefore energetically forced into narrow, string-like tubes. For a winding number $k$, the trapped flux is quantized, schematically as,

\begin{equation}
 \Phi_B = \oint A_\theta \, r \, d\theta = \frac{2\pi k}{n e_h} \, , \qquad k \in \mathbb{Z} \, ,
\end{equation}
so that the resulting objects are stable localized carriers of hidden magnetic flux.

In this way, the late-time symmetry breaking does not merely leave behind a bath of free hidden magnetic radiation. Rather, it reorganizes the relevant hidden energy into an extended network of flux-tube defects whose microscopic width is set by the inverse hidden Higgs and gauge masses, while their lengths may be macroscopic on cosmological scales. The detailed nonequilibrium dynamics of the transition, including defect formation, intercommutation, loop production, and the subsequent network evolution \cite{HindmarshKibble1995, VilenkinShellard1994}, are not required at the phenomenological level. For the present proof-of-concept analysis, it is sufficient to treat the magnetic flux network in coarse-grained form as a hidden flux-tube reservoir with an effective energy density $\rho_{\rm flux}$ and an effective equation-of-state parameter $w_{\rm flux}$.

Once such a reservoir is present, the next question is how it communicates with the vacuum-like sector associated with the intrinsic tension of space. In particular, if the hidden defect network can exchange energy with the dark-energy sector, then the total tension can no longer remain purely rigid at late times. This motivates a decomposition of the space tension into a constant background piece and a dynamical contribution, from which a time-dependent effective equation of state for dark energy naturally emerges.

\subsection{\centering Splitting the Tension Sector and the Effective Equation of State of Dark Energy}

If we assume that the energy carried by the magnetic flux tube defects produced due to the breaking of $U(1)$ gauge symmetry by the hidden Higgs field can be exchanged with the vacuum sector, then the spatial tension $T_s$ identified in Sec.~2 splits into a constant background piece and a dynamical transferred piece,

\begin{equation}
 T(t) = T_s + \delta T_s(t) \, , \qquad \dot T_s = 0 \, .
 \label{eq:Ts_split}
\end{equation}
Here $T_s$ denotes as before the rigid background contribution that survives after the ultraviolet physics has played its role. It is identified with the dominant constant part of the observed dark-energy scale as mentioned earlier. By contrast, $\delta T_s$ represents the late-time correction induced by energy exchange with the hidden flux-tube sector. Only $\delta T_s$ is allowed to participate in the exchange. This is physically important as the constant background $T_s$ is the baseline vacuum sector itself and is understood as a built-in background parameter inherited from underlying ultraviolet physics. In that sense it is rigid and cannot be treated as a reservoir that either drains energy into other sectors or receives energy back from them. It is simply the irreducible vacuum baseline that remains after the ultraviolet problem has already been resolved.

What can evolve is only an additional contribution deposited on top of this baseline by later hidden-sector dynamics. Therefore \(\delta T_s\) denotes the excess energy temporarily stored in the tension sector. Any subsequent relaxation, unloading, or transfer should then be understood as acting only on \(\delta T_s\), not on the underlying vacuum piece \(T_s\) itself. The late-time dynamics does not remove the intrinsic vacuum tension of space. It only perturbs the total tension away from its rigid baseline and may later allow that excess contribution to decay.

At the background level, we continue to identify the dark-energy density with the total space tension,

\begin{equation}
 \rho_{\rm DE}(t) \; \equiv \; T(t) \; = \; T_s + \delta T_s(t) \, .
 \label{eq:rhoDE_Ts}
\end{equation}
Since $\rho_{\rm DE}$ changes with time, or equivalently the scale factor $a$, its effective equation of state is also time dependent and reads (see Appendix A),

\begin{equation}
 w_{\rm eff}(a) = - 1 - \frac{1}{3} \frac{d\ln T}{d\ln a} = - 1 - \frac{1}{3\bigl(T_s + \delta T_s\bigr)} \frac{d\delta T_s}{d\ln a} \, .
 \label{eq:weff_general_deltaTs}
\end{equation}
One can therefore see that if $\delta T_s$ changes with the scale factor, then $w_{\rm eff}$ changes as well, giving a dynamical equation of state for dark energy.

Equation \eqref{eq:weff_general_deltaTs} makes the central point of the present framework explicit. Any late-time exchange that modifies $\delta T_s$ immediately induces a departure of $w_{\rm eff}$ from $-1$, even though the rigid background contribution $T_s$ itself remains untouched. In this sense, the running of dark energy is not attributed to the erosion of the vacuum baseline, but to the evolution of an additional tension component deposited on top of it by hidden-sector dynamics. What remains is to represent this exchange in a simple local effective description. This is the purpose of the next subsection.

\subsection{\centering Pedagogical Lagrangian for the coupled Tension and Flux Sectors}

The splitting introduced above is sufficient to define the background dark-energy sector and its effective equation of state, but it does not yet specify a local field-theoretic channel through which the hidden flux reservoir communicates with the dynamical part of the tension. To make that interaction more concrete, it is useful to introduce a simple pedagogical effective action in which the transferred part of the tension is encoded by a slowly varying scalar degree of freedom \cite{copeland2006, Amendola2000}. This provides a compact local framework for organizing the couplings between the tension sector, the hidden gauge field, and the hidden Higgs field.

To write a local effective action for a dynamical exchange, it is convenient to represent the transferred part of the tension sector by a slowly varying scalar degree of freedom $\sigma$ such that,

\begin{equation}
 T(\sigma) = T_s + \delta T_s(a, \sigma) \, .
 \label{eq:Ts_sigma}
\end{equation}
A simple effective action follows from augmenting equation \eqref{eq:hidden_U1_Zn_L} with a term to encode the dynamics of the $\sigma$ sector and replace $T_s$ with $T = T_s + \delta T_s$,

\begin{align}
 S = \int d^4x \, \sqrt{-g} \, \Bigg[- \Bigl(T_s & + \delta T_s (a, \sigma)\Bigr) - \frac{1}{4} Z(\sigma) \, F_{\mu\nu} \, F^{\mu\nu} \\ \nonumber
 & + |D_\mu \Phi|^2 - \frac{\lambda_\Phi}{4} \left( |\Phi|^2 - v_\Phi^2(\sigma) \right)^2  - \frac{K_\sigma}{2}\,(\partial^\mu \sigma \,  \partial_\mu \sigma) \Bigg] \, ,
 \label{eq:toy_action_coupled}
\end{align}
where the quantity $K_{\sigma}$ is a constant. Furthermore, $\delta T_s(\sigma)$ makes explicit that only the dynamical correction to the tension is carried by $\sigma$, while $T_s$ remains constant. The couplings $Z(\sigma)$ and $v_\Phi(\sigma)$ provide an effective communication channel between the hidden flux-tube sector and the tension sector.

As mentioned earlier, the present treatment is intentionally \textit{phenomenological}. We do not attempt to solve the full defect-formation and string-network dynamics from the field theory above. Instead, we move to a coarse-grained background description.

\subsection{\centering Coarse-grained Continuity Equations}

Let $\rho_{\rm flux}$ denote the coarse-grained energy density of the hidden
flux-tube network, and let its effective pressure be parameterized by,

\begin{equation}
 p_{\rm flux} = w_{\rm flux} \, \rho_{\rm flux} \, .
 \label{eq:wflux_def}
\end{equation}
For an almost frozen string-like reservoir one expects $w_{\rm flux}$ to be
negative, while for a more rapidly evolving network it may be closer to zero. In the present proof-of-concept treatment we leave $w_{\rm flux}$ as an effective parameter.

We now assume that the transferred part of the tension sector and the flux-tube reservoir exchange energy through a bidirectional channel. The corresponding continuity equations are taken to be (see Appendix B),

\begin{align}
 \dot{\rho}_{\rm flux} + 3 \, H\bigl(1 + w_{\rm flux}\bigr)\rho_{\rm flux} & = - \, Q \, , 
 \label{eq:rhoflux_cont}
 \\[1ex]
 \dot{\delta T_s} & = Q \, .
 \label{eq:deltaTs_cont}
\end{align}
where $Q$ is the amount of energy density exchanged per unit time. The exchange term is modeled phenomenologically as,

\begin{equation}
 Q = \Gamma H \left( \rho_{\rm flux} - \kappa \, T_s \right) \, , \qquad \Gamma \, > \, 0 \, , \qquad 0 \, \leq \, \kappa \, \leq \, 1 \, .
 \label{eq:Q_exchange}
\end{equation}
where $\Gamma$ is a dimensionless coupling parameter. This form captures the simplest balanced communication channel between the two dynamical sectors. We note that the phenomenological exchange law introduced above was written in its minimal form by taking the coupling parameter $\Gamma$ to be constant. However, once the transferred part of the tension sector is encoded by the scalar degree of freedom $\sigma$ as discussed in Sec.~3.3, it is natural to expect that the efficiency of the exchange channel itself may depend on the state of that sector. This expectation is reinforced by the fact that the same scalar already controls the couplings $Z(\sigma)$ and $v_\Phi(\sigma)$ in the local effective action. However, here we keep the analysis simple and treat $\Gamma$ as a constant. When the hidden flux-tube reservoir dominates over the stored dynamical correction to the space tension, that is when

\begin{equation}
 \rho_{\rm flux} \, > \, \kappa \, T_s \, ,
\end{equation}
one has $Q \, > \, 0$, so energy flows from the hidden defect sector into the space tension. By contrast, when

\begin{equation}
 \rho_{\rm flux} \, < \, \kappa \, T_s \, ,
\end{equation}
one has $Q \, < \, 0$, so the excess dynamical part of the tension sector relaxes back into the hidden flux sector. Importantly, the constant background piece $T_s$ never participates in this exchange.

Substituting \eqref{eq:Q_exchange} into \eqref{eq:rhoflux_cont} and
\eqref{eq:deltaTs_cont}, the coupled background system becomes,

\begin{align}
 \dot{\rho}_{\rm flux} + 3 H \bigl(1 + w_{\rm flux}\bigr)\rho_{\rm flux}
 & = - \, \Gamma H \left( \rho_{\rm flux} - \kappa \, T_s \right) \, , 
 \label{eq:rhoflux_final}
 \\[1ex]
 \dot{\delta T_s} & = \Gamma H \left( \rho_{\rm flux} - \kappa \, T_s \right) \, .
 \label{eq:deltaTs_final}
\end{align} 
If we further assume that the energy density of the flux-tube network is some fraction $\eta$ of the residual vacuum energy $T_s$, then,

\begin{equation}
 \rho_{\rm flux} = \eta(t) \, T_s \, , \qquad 0 \, \leq \, \eta(t) \, \leq \, 1
\end{equation}
which gives

\begin{align}
 \dot{\eta} + 3 H \bigl(1 + w_{\rm flux}\bigr) \eta & = - \Gamma H (\eta-\kappa) \, ,
 \label{eq:eta_t}
 \\[1ex]
 \dot{\delta T_s} & = \Gamma H T_s (\eta-\kappa) \, .
 \label{eq:deltaTs_t_eta}
\end{align}
Note that here we have assumed $\eta(t)$ to be a time dependent function because the energy density of the flux-tube network will decrease as a function of time due to cosmic expansion.

Writing the evolution in terms of the e-folding variable,

\begin{equation}
 a = e^N \, , \qquad N \equiv \ln a \, , \qquad \frac{d}{dN} = \frac{1}{H}\frac{d}{dt} \, ,
 \label{eq:N_def}
\end{equation}
the system becomes,

\begin{align}
 \frac{d\eta}{dN} + 3 \bigl(1 + w_{\rm flux}\bigr) \eta & = - \Gamma \, (\eta - \kappa)
 \label{eq:eta_N}
 \\[1ex]
 \frac{d \, \delta T_s}{dN} & = \Gamma \, T_s(\eta - \kappa) \, .
 \label{eq:deltaTs_N}
\end{align}
Equation \eqref{eq:eta_N} further gives,

\begin{equation}
 \frac{d\eta}{dN} = - \Bigl[3 \bigl(1 + w_{\rm flux}\bigr) + \Gamma\Bigr] \eta + \Gamma \, \kappa \, ,
\end{equation}
If \(w_{\rm flux}\), \(\Gamma\), and \(\kappa\) are treated as constants, the solution for \(\eta(a)\) is,

\begin{equation}
 \eta(a) = \eta_\infty + \bigl(\eta_i - \eta_\infty\bigr) \left(\frac{a}{a_i}\right)^{-m} \, ,
 \label{eq:eta_solution}
\end{equation}
with,

\begin{equation}
 m = 3\bigl(1 + w_{\rm flux}\bigr) + \Gamma \, , \qquad \eta_\infty = \frac{\Gamma \, \kappa}{m} \, .
 \label{eq:m_eta_inf}
\end{equation}
In equation \eqref{eq:eta_solution}, the term $a_i$ and $\eta_i$ denote the initial value of both, the scale factor and the hidden flux fraction, at that point in time when the hidden Higgs field condenses, thereby breaking the symmetry of the DBI gauge field from $U(1) \to \mathbb{Z}_n$. We further write,

\begin{equation}
 m \; = \; 3 \bigl(1 + w_{\rm flux}\bigr) + \Gamma \; = \; n + \Gamma \, , \qquad n \equiv 3 \bigl(1 + w_{\rm flux}\bigr)
 \label{w flux and n relation}
\end{equation} 
Now, using equation \eqref{eq:deltaTs_N} in \eqref{eq:weff_general_deltaTs}, we obtain,

\begin{equation}
 w_{\rm eff}(a) = - 1 - \frac{\Gamma \, T_s}{3\bigl[T_s + \delta T_s(a)\bigr]} \bigl(\eta(a) - \kappa\bigr) \, .
 \label{eq:weff_eta}
\end{equation}
This makes the phantom-divide crossing transparent. If the hidden flux-tube sector is initially large enough such that,

\begin{equation}
 \eta(a) > \kappa \, ,
 \label{eq:eta_gt_kappa}
\end{equation}
then,

\begin{equation}
 \frac{d\,\delta T_s}{dN} > 0 \qquad \Longrightarrow \qquad w_{\rm eff}(a) < -1 \, .
 \label{eq:phantom_case}
\end{equation}
If,

\begin{equation}
 \eta(a) = \kappa \, ,
 \label{eq:eta_eq_kappa}
\end{equation}
then,

\begin{equation}
 \frac{d\,\delta T_s}{dN} = 0 \qquad \Longrightarrow \qquad w_{\rm eff}(a) = - 1 \, .
 \label{eq:crossing_case}
\end{equation}
At later times, the hidden defect reservoir redshifts and depletes. Once the
balance reverses, one has,

\begin{equation}
 \eta(a) < \kappa \, ,
 \label{eq:eta_lt_kappa}
\end{equation}
then,

\begin{equation}
 \frac{d\,\delta T_s}{dN} < 0 \qquad \Longrightarrow \qquad w_{\rm eff}(a) > -1 \, .
 \label{eq:quintessence_case}
\end{equation}
The model therefore admits, at the level of a simple coarse-grained effective description, a transient crossing of the phantom divide. In this way the hidden flux-tube reservoir generated by the late-time $U(1)_h \to \mathbb{Z}_n$ transition provides a physically transparent proof of concept for a mild low-redshift running of the effective dark-energy sector.

The coarse-grained system derived above yields an explicit late-time evolution for $\eta(a)$, $\delta T_s(a)$, and hence for the effective dark-energy equation of state $w_{\rm eff}(a)$. The natural next step is to ask whether this predicted running is phenomenologically relevant. Rather than confronting the full cosmological likelihood directly, we adopt a simpler benchmark strategy and compare the model evolution with the effective CPL description preferred by current late-time data. This allows us to assess, in a transparent way, how closely the present running-tension scenario can approach the observationally favored region in the $(w_0,w_a)$ plane.

\section{\centering Observational Benchmarks and Results}

In this section we summarize the numerical and phenomenological results of the flux-tube running-tension model. The purpose is not to perform a full Monte Carlo likelihood analysis, but to compare the model with compressed late-time dark-energy benchmarks and to extract the effective parameters that characterize the resulting low-redshift equation-of-state evolution.

The coarse-grained system developed in Section 3 yields a late-time prediction for the hidden flux fraction $\eta(a)$, the dynamical tension shift $\delta T_s(a)$, and therefore the effective dark-energy equation of state $w_{\rm eff}(a)$. The purpose of the present section is to test this prediction against a simple observational benchmark suggested by current late-time data. More specifically, we ask whether the flux-tube driven running of the tension sector can reproduce the kind of mild low-redshift evolution commonly summarized in the Chevallier--Polarski--Linder, or CPL plane.

Our aim here is diagnostic rather than inferential. We do not attempt a full end-to-end cosmological likelihood analysis of the hidden-sector model. Instead, we use a compressed benchmark built from the public $w_0w_a$ posterior and compare the resulting CPL target with the explicit curve $w_{\rm eff}(z)$ predicted by the running-tension model. This is sufficient for the proof-of-concept purpose of the present work, which is to determine whether a late hidden defect reservoir can induce a phenomenologically relevant departure from a strictly constant dark-energy sector. A conceptually similar compressed-benchmark comparison, applied to a different dynamical dark-energy model based on spatial phonons, was also adopted in \cite{KhanSpatialPhonons2025}.

\subsection{\centering Benchmark Data and Effective CPL Target}

As the observational reference for the present analysis, we use the publicly released posterior chains for the flat \(w_0w_a\)CDM extension of \(\Lambda\)CDM from the DESI DR1 BAO value-added cosmology products. Specifically, we use the dataset combination DESI DR1 BAO + Pantheon+ + Planck 2018. Since the purpose of this section is a diagnostic comparison rather than a full likelihood analysis, we do not propagate the full posterior samples. Instead, we compress the relevant information into a two-parameter Gaussian benchmark in the CPL plane.

Let
\begin{equation}
 \hat{\theta} =
 \begin{pmatrix}
  \hat{w}_0 \\
  \hat{w}_a
 \end{pmatrix}
 \label{eq:thetahat_def}
\end{equation}
denote the weighted posterior mean of the released chains, and let \(C\) denote the corresponding weighted covariance matrix. The compressed benchmark used below is therefore specified by the pair \((\hat{\theta},C)\). This compression retains the central location, variance, and covariance of the posterior in the \((w_0,w_a)\) plane, but it does not retain possible non-Gaussian features of the full posterior or correlations with other cosmological parameters.

The weighted mean vector and covariance matrix were extracted from the released posterior samples using the custom Python script \texttt{w\_0\_w\_a\_python.py}, which is available at Ref.~\cite{KhanMGKSPDataset_2}. Applying this script to the DESI DR1 BAO + Pantheon+ + Planck 2018 chains gives

\begin{equation}
 \hat{\theta} = 
  \begin{pmatrix}
  \hat{w}_0 \\
  \hat{w}_a
 \end{pmatrix} =
 \begin{pmatrix}
  -0.828209 \\
  -0.744750
 \end{pmatrix},
 \qquad
 C =
 \begin{pmatrix}
  0.004112 & -0.016695 \\
  -0.016695 & 0.085227
 \end{pmatrix}.
 \label{eq:cpl_benchmark_numbers}
\end{equation}
Equation \eqref{eq:cpl_benchmark_numbers} is the only observational input required for the analysis that follows. The benchmark is therefore intentionally lightweight. It does not retain non-Gaussian structure in the full posterior, nor does it track correlations with the other cosmological parameters that would enter a full global fit. That loss of information is acceptable for the limited purpose of the present section, which is only to assess whether the flux-tube mechanism can generate a low-redshift evolution with roughly the right shape and amplitude.

For additional perspective, it is useful to compare the result obtained above for $w_0$ and $w_a$ from a rather simple analysis of the \texttt{cobaya} chains with the analyses presented by Chudaykin and Kunz \cite{ChudaykinKunz2024}. In their baseline $w_0w_a$CDM analysis of the CMB + DESI + SN combination, they found,

\begin{equation}
 w_0 = -0.821 \pm 0.065, \qquad w_a = -0.76^{+0.31}_{-0.26},
\end{equation}
which is very close to our quoted value in \eqref{eq:cpl_benchmark_numbers}. 

The CPL trajectory is defined as \cite{ChevallierPolarski2001, Linder2003},

\begin{equation}
 w_{\rm CPL}(z) = w_0 + w_a \frac{z}{1 + z}.
 \label{eq:cpl_form_section4}
\end{equation}
Substituting the benchmark mean values from equation \eqref{eq:cpl_benchmark_numbers} in the relation above gives the corresponding DESI-CPL benchmark trajectory as,

\begin{equation}
 w_{\rm CPL}^{\rm DESI}(z) = \hat{w}_0 + \hat{w}_a \, \frac{z}{1 + z} = - 0.828209 - 0.744750 \left(\frac{z}{1 + z}\right).
 \label{DESI-CPL-w}
\end{equation}
The curve generated from \(w_{\rm CPL}^{\rm DESI}(z)\) will serve as the target curve against which the equation of state \(w_{\rm eff}^{\rm model}(z)\) obtained from the running-tension model is tested.

Although the observational benchmark is expressed in the CPL plane, the quantity predicted directly by the present model is not a fundamental CPL pair \((w_0, \, w_a)\), but the full redshift-dependent effective equation of state \(w_{\rm eff}(a)\), given in Eq.~\eqref{eq:weff_general_deltaTs}. Recall that this equation of state is generated by the running of the total space tension,

\begin{equation}
 T(a) = T_s + \delta T_s(a).
\end{equation}
Accordingly, we first compare the two descriptions at the level of the equation-of-state curves themselves. That is, we compare the model prediction \(w_{\rm eff}^{\rm model}(z)\) with the compressed DESI-CPL benchmark curve \(w_{\rm CPL}^{\rm DESI}(z)\), defined in Eq.~\eqref{DESI-CPL-w}.

After this direct curve-level comparison, we project the model prediction \(w_{\rm eff}^{\rm model}(z)\) onto the CPL form in Eq.~\eqref{eq:cpl_form_section4}. This gives an effective running tension model CPL pair,

\begin{equation}
 (w_0^{\rm mod}, \, w_a^{\rm mod}) \, ,
\end{equation}
which can be compared with the DESI DR1 BAO + Pantheon+ + Planck 2018 benchmark values \((\hat{w}_0, \, \hat{w}_a)\) given in Eq.~\eqref{eq:cpl_benchmark_numbers}. This projected pair should not be interpreted as a set of fundamental parameters of the model. It is only a convenient two-parameter summary of the running-tension curve over the redshift range considered.

The projection is performed over the interval,

\begin{equation}
 0 \leq z \leq 1.6,
\end{equation}
using an ordinary least-squares fit onto the CPL basis,

\begin{equation}
 \left\{1, \, \frac{z}{1 + z}\right\}.
\end{equation}
In practice, the fit is evaluated on the same uniform scale-factor grid used for the pointwise comparison, with \(z=1/a-1\). We choose this interval because it captures the low-redshift region relevant for the present comparison and covers the range over which the DESI-motivated CPL benchmark is most useful for diagnosing late-time dark-energy evolution.

In summary, the comparison performed in this section involves two distinct objects.  On the observational side, the quantities \(\hat{w}_0\) and \(\hat{w}_a\) are obtained by applying a Gaussian compression to the publicly released DESI DR1 BAO + Pantheon+ + Planck 2018 posterior chains in the flat \(w_0w_a\)CDM model. Substituting these values into the CPL parametrization,

\begin{equation}
 w_{\rm CPL}^{\rm DESI}(z) = \hat{w}_0 + \hat{w}_a\frac{z}{1 + z} ,
\end{equation}
gives a compressed observational benchmark curve for the dark-energy equation of state as a function of redshift obtained from DESI studies.

On the model side, the running-tension scenario does not directly predict a fundamental CPL pair \((w_0, \, w_a)\). Instead, it predicts the full effective equation-of-state curve,

\begin{equation}
 w_{\rm eff}^{\rm model}(a) = - 1 - \frac{1}{3}\frac{d\ln T(a)}{d\ln a} \, , \qquad T(a) = T_s + \delta T_s(a).
\end{equation}
This curve can be rewritten as a function of redshift using \(a = (1 + z)^{-1}\). This gives

\begin{equation}
 w_{\rm eff}^{\rm model}(z) = - 1 + \frac{1 + z}{3} \, \frac{d\ln T(z)}{dz} \, .
 \label{eq:weff_z_logT_main}
\end{equation}
Using the definitions of $\eta$, $\kappa$, $\Gamma$, and $m$ as presented in Sec.~3.4, the above equation can be written as (see Appendix C),

\begin{equation}
\begin{aligned}
 w_{\rm eff}^{\rm model}(z) = -1 -
 \frac{\Gamma\left[\eta_\infty + (\eta_i-\eta_\infty)\left[a_i(1+z)\right]^m - \kappa\right]}
 {3\left[1 + \Gamma\left\{(\kappa-\eta_\infty)\ln\left[a_i(1+z)\right]
 + \dfrac{\eta_i-\eta_\infty}{m}\left(1-\left[a_i(1+z)\right]^m\right)\right\}\right]} \, .
\end{aligned}
 \label{eq:weff_model_z_explicit_main}
\end{equation}
The first comparison is therefore a direct curve-level comparison between \(w_{\rm CPL}^{\rm DESI}(z)\) and \(w_{\rm eff}^{\rm model}(z)\).

For a second, more compact comparison, we project the model curve \(w_{\rm eff}^{\rm model}(z)\) onto the CPL basis,

\begin{equation}
 \left\{1, \, \frac{z}{1 + z}\right\}
\end{equation}
over the interval \(0\leq z\leq 1.6\). This ordinary least-squares projection gives an effective model pair,

\begin{equation}
 (w_0^{\rm mod}, \, w_a^{\rm mod}),
\end{equation}
defined by,

\begin{equation}
 w_{\rm eff}^{\rm model}(z) \; \simeq \; w_0^{\rm mod} + w_a^{\rm mod} \, \frac{z}{1 + z}.
\end{equation}
The observational pair \((\hat{w}_0, \hat{w}_a)\) can then be compared with the projected model pair \((w_0^{\rm mod}, w_a^{\rm mod})\), while the full functions \(w_{\rm CPL}^{\rm DESI}(z)\) and \(w_{\rm eff}^{\rm model}(z)\) can be compared directly at the curve level. Thus the comparison is performed in two complementary ways: first through the full redshift-dependent equation-of-state curves, and second through their corresponding effective locations in the CPL plane.

\subsection{\centering Pointwise Fit of the Flux-Tube Running Tension Model to the DESI-CPL Curve}

We now test the flux-tube running-tension ansatz directly against the DESI-motivated effective CPL benchmark constructed in Sec.~4.1. At the background level, the dark-energy density in the present model is identified with the total effective space tension,

\begin{equation}
 \rho_{\rm DE}(a) = T(a) = T_s + \delta T_s(a).
 \label{eq:rhoDE_running_tension_sec42}
\end{equation}
The corresponding equation of state is the running-tension quantity,

\begin{equation}
 w_{\rm eff}^{\rm model}(a) = - 1 - \frac{1}{3} \, \frac{d\ln T(a)}{d\ln a} \, .
 \label{eq:weff_model_sec42}
\end{equation}
This is compared with the compressed DESI-CPL benchmark,

\begin{equation}
 w_{\rm CPL}^{\rm DESI}(a) = \hat{w}_0 + \hat{w}_a(1 - a), \qquad (\hat{w}_0, \hat{w}_a) = (-0.828209, -0.744750) ,
 \label{eq:desi_cpl_a_sec42}
\end{equation}
or equivalently Eq.~\eqref{DESI-CPL-w}.

The comparison is performed over the redshift interval,

\begin{equation}
 0 \leq z \leq 1.6,
\end{equation}
which corresponds to the scale-factor interval,

\begin{equation}
 a_{\min} \leq a \leq 1, \qquad a_{\min} = \frac{1}{1 + 1.6} \simeq 0.384615.
 \label{eq:amin_sec42}
\end{equation}
The hidden transition scale factor \(a_i\) is allowed to lie within this interval. For epochs before the hidden transition, \(a<a_i\), the hidden flux-tube reservoir has not yet formed and we take,

\begin{equation}
 w_{\rm eff}^{\rm model}(a) = - 1. 
\end{equation}
For \(a \geq a_i\), we use the analytic solution of the coarse-grained exchange equations derived in Sec.~3.4, with the initial condition,

\begin{equation}
 \delta T_s(a_i) = 0.
\end{equation}
The restricted scan is performed over,

\begin{equation}
 n = 3(1 + w_{\rm flux}) \in \{2, 3, 4\}, \qquad 0 \leq \eta_i \leq 1 , \qquad 0 \leq \kappa \leq 1, \qquad \Gamma \geq 0.
 \label{eq:scan_ranges_sec42}
\end{equation}
These restrictions have a simple physical interpretation. The parameter \(n\) controls the dilution of the hidden flux reservoir,

\begin{equation}
 n = 3(1 + w_{\rm flux}) .
\end{equation}
The choice \(n = 2\) corresponds to a string-like flux-tube network with \(w_{\rm flux} = -1/3\), \(n = 3\) corresponds to matter-like dilution with \(w_{\rm flux} = 0\), and \(n = 4\) corresponds to radiation-like dilution with \(w_{\rm flux} = 1/3\). Values \(n = 0\) or \(n = 1\) would correspond to \(w_{\rm flux} = -1\) or \(w_{\rm flux} = -2/3\), respectively, which are too vacuum-like or domain-wall-like for the flux-tube reservoir considered here. Values \(n > 4\) would require \(w_{\rm flux} > 1/3\), namely a reservoir redshifting faster than radiation, which is not expected for the coarse-grained hidden magnetic flux-tube, matter-like, or radiation-like regimes relevant to the present proof-of-concept model. We therefore restrict to the physically motivated set \(n = 2, 3, 4\).

The parameter \(\eta_i\) is the initial hidden flux fraction at the transition as mentioned previously in Sec.~3.4,

\begin{equation}
 \rho_{\rm flux}(a_i) = \eta_i T_s.
\end{equation}
Since it is a fractional energy density in the restricted minimal model, we impose \(0 \leq \eta_i \leq 1\). Larger values are not logically impossible, but they would correspond to a hidden reservoir initially exceeding the residual vacuum-tension scale and are therefore left outside the conservative benchmark scan. The parameter \(\kappa\) defines the balance value in the exchange term,

\begin{equation}
 Q = \Gamma H(\rho_{\rm flux} - \kappa T_s).
\end{equation}
Thus \(\kappa T_s\) is the flux density at which the direction of energy exchange reverses. Taking \(0 \leq \kappa \leq 1\) means that this balance density is also restricted to lie between zero and the residual tension scale. Finally, \(\Gamma\) is taken to be non-negative because it represents the strength of the exchange channel. With the sign convention above, \(\Gamma > 0\) means that a flux reservoir above the balance value feeds the dynamical tension correction, while a reservoir below the balance value allows the excess tension correction to relax back into the hidden sector. The limiting case \(\Gamma = 0\) gives no exchange.

To quantify the agreement with the DESI-CPL benchmark, we sample the interval \(a \in [a_{\min}, 1]\) using \(N = 100\) uniformly spaced points \(a_j\), with,

\begin{equation}
 z_j = \frac{1}{a_j} - 1 .
\end{equation}
The pointwise root-mean-square error is defined as,

\begin{equation}
 {\rm RMSE} = \left[ \frac{1}{N} \sum_{j = 1}^{N} \left( w_{\rm eff}^{\rm model}(a_j) - w_{\rm CPL}^{\rm DESI}(a_j) \right)^2 \right]^{1/2}.
 \label{eq:rmse_sec42}
\end{equation}
For each parameter point, we also compute an effective CPL projection of the model curve by fitting,

\begin{equation}
 w_{\rm eff}^{\rm model}(z_j) \simeq w_0^{\rm mod} + w_a^{\rm mod} \, \frac{z_j}{1 + z_j}
\end{equation}
on the same grid. The resulting pair \((w_0^{\rm mod}, w_a^{\rm mod})\) is not a fundamental parameter pair of the model. It is only a compact CPL summary of the running-tension curve over the redshift interval \(0 \leq z \leq 1.6\).

The initial coarse scan used the uniform grid,

\begin{equation}
 N_{a_i} = 31 , \qquad N_{\eta_i} = 31, \qquad N_{\Gamma} = 61, \qquad N_{\kappa} = 31,
\end{equation}
which corresponds to
\begin{equation}
 31 \times 31 \times 61 \times 31 = 1{,}815{,}771
\end{equation}
models for each fixed value of \(n\), or \(5{,}447{,}313\) models in total for \(n = 2 , 3, 4\). The Python script used for the scan is available as \texttt{scan\_flux\_fit.py} at Ref.~\cite{KhanMGKSPDataset_2}. The best region of the coarse scan was found near \(n = 2\), \(a_i = a_{\min}\), \(\eta_i \simeq 1\), \(\Gamma \simeq 1.45\), and \(\kappa \simeq 0.17\).

We then performed a local refinement around this region. In the refined scan we fixed,

\begin{equation}
 n = 2, \qquad a_i = a_{\min} = 0.384615 ,
\end{equation}
and scanned more densely over,

\begin{equation}
 \eta_i \in [0.90, 1.00], \qquad \Gamma \in [1.20, 1.70], \qquad \kappa \in [0.05, 0.30],
\end{equation}
using a \(41 \times 101 \times 101\) grid. The final refined best-fit point is given in Table~\ref{tab:flux_fit_final}. This table contains the final parameter values used in the subsequent plots and comparisons.

\begin{table}[htbp]
 \centering
 \begin{tabular}{l c l}
  \hline
  Quantity & Value & Interpretation \\
  \hline
  \(n\) & \(2\) & String-like flux dilution index \\
  \(w_{\rm flux}\) & \(-1/3\) & Effective equation of state of flux-tube reservoir \\
  \(a_i\) & \(0.384615\) & Hidden transition scale factor \\
  \(z_i\) & \(1.6000\) & Hidden transition redshift \\
  \(\eta_i\) & \(1.0000\) & Initial hidden flux fraction \\
  \(\Gamma\) & \(1.45\) & Dimensionless exchange strength \\
  \(\kappa\) & \(0.1525\) & Balance parameter for flux--tension exchange \\
  \(m = n + \Gamma\) & \(3.45\) & Effective depletion exponent \\
  \(\eta_\infty = \Gamma\kappa/m\) & \(0.06409\) & Asymptotic hidden flux fraction \\
  \(z_\times\) & \(0.3120\) & Redshift of phantom-divide crossing \\
  \(w_{\rm eff}^{\rm model}(z = 1.6)\) & \(-1.4096\) & Model value at transition redshift \\
  \(w_{\rm eff}^{\rm model}(z = 0)\) & \(-0.9793\) & Present-day model value \\
  \(w_0^{\rm mod}\) & \(-0.9153\) & Effective CPL projection of model curve \\
  \(w_a^{\rm mod}\) & \(-0.4572\) & Effective CPL projection of model curve \\
  RMSE & \(0.07463\) & Pointwise deviation from DESI-CPL benchmark \\
  \hline
 \end{tabular}
 \caption{Final refined best-fit point of the flux-tube running-tension model against the DESI-motivated effective CPL benchmark. The entries \(n, a_i, \eta_i, \Gamma, \kappa\) are model parameters, while \(z_\times\), \(w_{\rm eff}^{\rm model}(z)\), \((w_0^{\rm mod}, w_a^{\rm mod})\), and RMSE are derived outputs of the present model.}
 \label{tab:flux_fit_final}
\end{table}
\FloatBarrier \noindent The refined point has,

\begin{equation}
 {\rm RMSE} = 0.07463.
\end{equation}
The corresponding model projection on the CPL basis is,

\begin{equation}
 (w_0^{\rm mod}, w_a^{\rm mod}) = (-0.9153, -0.4572).
 \label{eq:model_cpl_projection_final}
\end{equation}
These values should be compared with the compressed DESI-CPL benchmark,

\begin{equation}
 (\hat{w}_0, \hat{w}_a) = (-0.828209, -0.744750).
\end{equation}
The model therefore does not exactly reproduce the compressed DESI-CPL target. Nevertheless, it generates a low-redshift running of the correct qualitative type and comparable amplitude. In particular, it evolves from a phantom-like value shortly after the hidden transition toward a value close to \(-1\) today, crossing the phantom divide at,

\begin{equation}
 z_\times \simeq 0.312.
\end{equation}
This crossing is not inserted by hand. It occurs when the hidden flux fraction passes through the balance value,

\begin{equation}
 \eta(a_\times) = \kappa.
\end{equation}
Thus the numerical fit supports the main physical point of the construction: a late hidden flux-tube reservoir can induce a mild running of the residual vacuum-like tension of space while leaving the rigid background tension \(T_s\) intact.

Two qualitative features are noteworthy. First, the fit places the hidden transition at the earliest redshift allowed within the comparison window,
\begin{equation}
 a_i = a_{\min} \simeq 0.384615, \qquad z_i = 1.6 .
\end{equation}
Thus the preferred transition is not an early-universe event in the usual sense. It occurs well after recombination, during the low-redshift matter-dominated era, precisely in the range where a mild evolution of the dark-energy sector could become phenomenologically relevant. In the language of the present model, this means that the hidden Higgs condensation is favored to occur as a genuinely late-time transition.

Such a late onset should not be viewed as implausible merely because it occurs at low redshift. Since the relevant sector is hidden, its symmetry-breaking history need not be tied to the thermal history of the visible Standard Model. Delayed condensation could arise, for example, through a temperature-dependent hidden effective potential, through couplings to a slowly varying background degree of freedom such as the tension-sector scalar, or through radiative and threshold effects that postpone symmetry breaking until late times. We do not attempt to construct such a microscopic mechanism here. Nevertheless, the hidden nature of the sector makes late-time condensation a reasonable phenomenological possibility rather than an exceptional assumption. A detailed realization of this transition is therefore an important direction for future work.

Second, the fit drives the initial hidden flux fraction to its maximal value,

\begin{equation}
 \eta_i = 1.
\end{equation}
Within the minimal single-reservoir ansatz, this indicates that the model prefers a strongly populated hidden defect sector in order to reproduce the DESI-motivated low-redshift running. At the same time, the resulting effective CPL projection remains visibly displaced from the benchmark target,

\begin{equation}
 (w_0^{\rm mod}, w_a^{\rm mod}) = (-0.91529, -0.45716), \qquad (\hat w_0, \hat w_a) = (-0.828209, -0.744750).
\end{equation}
The mismatch is therefore real, but not large enough to invalidate the mechanism as a proof of concept. Rather, it shows that the minimal one-reservoir model captures part of the desired low-redshift behavior while still leaving room for improvement. It is possible that a more complete treatment of the hidden-sector dynamics, defect-network evolution, and exchange channel with the tension sector could reduce the remaining gap to the observational benchmark.

Furthermore, substituting \(n = 2\) into

\begin{equation}
 n = 3(1 + w_{\rm flux})
\end{equation}
gives,

\begin{equation}
 w_{\rm flux} = \frac{n}{3} - 1 = -\frac{1}{3}.
\end{equation}
Thus the best-fit hidden flux sector behaves as an effectively string-like network of line-like defects. This is physically consistent with the picture developed in Sec.~3, where the late-time broken phase reorganizes hidden magnetic energy into flux tubes rather than into a bath of freely propagating relativistic quanta. A component with \(w_{\rm flux}=-1/3\) dilutes more slowly than radiation and behaves in a curvature-like manner at the background level. In the present framework this persistence is important: it allows the hidden defect reservoir to survive long enough, and with sufficient strength, to exchange energy with the dynamical part of the space tension and thereby induce a nontrivial evolution of \(w_{\rm eff}(a)\).

Table~\ref{tab:desi_model_comparison} compares the compressed DESI-CPL benchmark curve with the direct equation-of-state curve predicted by the running-tension model. The DESI quantity is,

\begin{equation}
 w_{\rm CPL}^{\rm DESI}(z) = \hat w_0 + \hat w_a \, \frac{z}{1 + z} ,
\end{equation}
whereas the model quantity is computed directly from

\begin{equation}
 w_{\rm eff}^{\rm model}(a) = - 1 -\frac{1}{3} \, \frac{d\ln T(a)}{d\ln a}, \qquad T(a) = T_s + \delta T_s(a).
\end{equation}
Therefore the table is a curve-level comparison, not merely a comparison between two CPL projections.

\begin{figure}[htbp]
 \centering
 \includegraphics[width=0.82\textwidth]{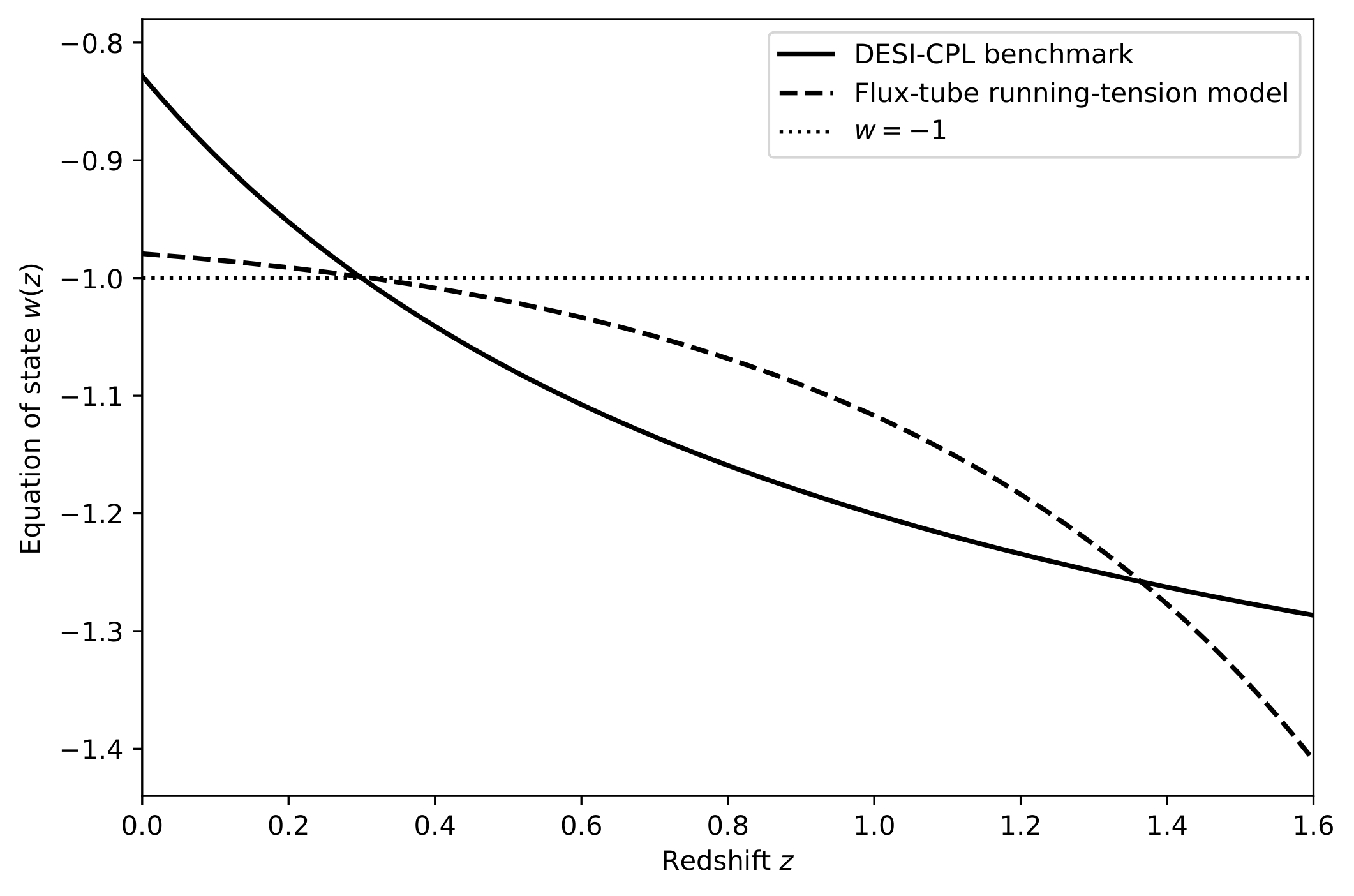}
 \caption{\footnotesize{Comparison between the DESI-motivated CPL benchmark equation of state and the best-fit flux-tube running-tension model over the range \(0 \le z \le 1.6\). The solid curve shows the CPL form \(w_{\rm CPL}(a) = w_0 + w_a(1 - a)\) with the benchmark parameters \((w_0,w_a) = (-0.828209, -0.744750)\), while the dashed curve shows the model prediction obtained from the running-tension relation \(w_{\rm eff}(a) = - 1 - \dfrac{1}{3} \dfrac{d\ln T}{d\ln a}\), with \(T(a) = T_s + \delta T_s(a)\). The model curve is evaluated using the best-fit restricted-\(n\) point from the 100-point scan, namely \(n = 2\), \(a_i = 1/2.6\), \(\eta_i = 1\), \(\Gamma = 1.45\), and \(\kappa = 0.1525\). The figure illustrates that, although the model captures a qualitatively evolving low-redshift equation of state, it does not reproduce the DESI-CPL curve pointwise across the full fitting interval. The dotted horizontal line denotes \(w=-1\).}}
 \label{fig:desi_model_comparison_curve}
\end{figure}

\begin{table}[htbp]
 \centering
 \begin{tabular}{c c c c c}
  \hline
  \(a\) & \(z\) & \(w_{\rm CPL}^{\rm DESI}\) & \(w_{\rm eff}^{\rm model}\) & \(\Delta w\) \\
  \hline
  0.384615 & 1.600000 & -1.286517 & -1.409625 & -0.123108 \\
  0.459207 & 1.177665 & -1.230964 & -1.175142 & \;\;0.055823 \\
  0.533800 & 0.873362 & -1.175412 & -1.084349 & \;\;0.091063 \\
  0.608392 & 0.643678 & -1.119859 & -1.040081 & \;\;0.079778 \\
  0.682984 & 0.464164 & -1.064307 & -1.015535 & \;\;0.048772 \\
  0.757576 & 0.320000 & -1.008754 & -1.000711 & \;\;0.008043 \\
  0.832168 & 0.201681 & -0.953202 & -0.991182 & -0.037980 \\
  0.906760 & 0.102828 & -0.897650 & -0.984749 & -0.087100 \\
  1.000000 & 0.000000 & -0.828209 & -0.979315 & -0.151106 \\
  \hline
 \end{tabular}
 \caption{Pointwise comparison between the compressed DESI-CPL benchmark and the best-fit flux-tube running-tension model over \(0 \leq z \leq 1.6\). The column \(w_{\rm CPL}^{\rm DESI}\) is obtained from the compressed DESI DR1 BAO + Pantheon+ + Planck 2018 CPL benchmark, while \(w_{\rm eff}^{\rm model}\) is the direct prediction of the running-tension model with \(n = 2\), \(a_i = 1/2.6\), \(\eta_i = 1\), \(\Gamma = 1.45\), and \(\kappa = 0.1525\). Here \(\Delta w \equiv w_{\rm eff}^{\rm model} - w_{\rm CPL}^{\rm DESI}\).}
 \label{tab:desi_model_comparison}
\end{table}
\FloatBarrier
\noindent Finally, Fig.~\ref{fig:desi_model_comparison_curve} shows the direct curve-level comparison between the DESI-CPL benchmark and the running-tension model. This plot complements the CPL-plane comparison. The CPL-plane plot compresses the model into two numbers, \((w_0^{\rm mod}, \, w_a^{\rm mod})\), while the curve-level plot shows the full redshift dependence of \(w_{\rm eff}^{\rm model}(z)\). The model does not exactly trace the DESI-CPL curve over the entire redshift interval, but it does reproduce the main qualitative feature: a low-redshift departure from \(w = -1\) with a dynamically generated phantom crossing.

\FloatBarrier

\subsection{\centering Pantheon+ Distance-Modulus Residual Diagnostic}

The comparison above tests the running-tension model at the level of an effective equation-of-state curve. It is also useful to examine the same background evolution at the level of Type Ia supernova distances. For this purpose we use the Pantheon+ compilation as an additional diagnostic. Unlike the DESI-CPL benchmark, Pantheon+ does not directly provide a measured function \(w(z)\). It constrains the expansion history through luminosity distances. Therefore, the appropriate comparison is made at the level of the distance modulus rather than by constructing a separate Pantheon-only equation-of-state curve.

For a spatially flat late-time background, and neglecting radiation over the redshift range relevant for the present comparison, the Hubble function of the running-tension model may be written as,

\begin{equation}
 E_{\rm mod}^2(z) \equiv \frac{H_{\rm mod}^2(z)}{H_0^2} = \Omega_{m0}(1 + z)^3 + \Omega_{{\rm DE}, 0} \frac{T(z)}{T(0)} \, .
 \label{eq:E_model_Tz}
\end{equation}
Here \(E_{\rm mod}(z)\) is the dimensionless expansion rate of the running-tension model. It is defined by normalizing the model Hubble parameter \(H_{\rm mod}(z)\) by the present-day Hubble constant \(H_0\). The quantity \(\Omega_{m0}\) denotes the present-day matter density fraction, and the factor \((1 + z)^3\) is the usual dilution of nonrelativistic matter with redshift. The quantity \(\Omega_{{\rm DE}, 0}\) denotes the present-day dark-energy density fraction. In the present model the dark-energy density is identified with the total effective space tension \(T(z)\), so its redshift evolution is encoded by the ratio \(T(z)/T(0)\). This ratio is normalized to unity today. In the special case where the tension is exactly constant, \(T(z) = T(0)\), the expression reduces to the standard flat \(\Lambda\)CDM form, neglecting radiation.

Now, since,

\begin{equation}
 T(z) = T_s \left[1 + \Delta(z)\right], \qquad \Delta(z) \equiv \frac{\delta T_s(z)}{T_s},
\end{equation}
this becomes,

\begin{equation}
 E_{\rm mod}^2(z) = \Omega_{m0}(1 + z)^3 + \Omega_{{\rm DE}, 0} \frac{1 + \Delta(z)}{1 + \Delta(0)}.
 \label{eq:E_model_Delta_z}
\end{equation}
The explicit expression for \(\Delta(z)\) is derived in Appendix~C.

The corresponding luminosity distance is,

\begin{equation}
 d_L^{\rm mod}(z) = (1+z)\frac{c}{H_0} \int_0^z \frac{dz'}{E_{\rm mod}(z')},
 \label{eq:luminosity_distance_model}
\end{equation}
and the distance modulus is,

\begin{equation}
 \mu_{\rm mod}(z) = 5 \log_{10} \left( \frac{d_L^{\rm mod}(z)}{{\rm Mpc}}
 \right) + 25.
 \label{eq:distance_modulus_model}
\end{equation}
The distance modulus is used because Type Ia supernovae constrain the luminosity-distance relation rather than the equation of state directly. The equation of state affects the Hubble rate \(H(z)\), which then affects the luminosity distance through an integral over \(1/H(z)\). The observable quantity is usually expressed as the distance modulus $\mu$ given above. It is therefore more appropriate to compare the running-tension model with Pantheon+ at the level of \(\mu(z)\), or more specifically through the residual,

\[
\Delta\mu(z)=\mu_{\rm mod}(z)-\mu_{\Lambda{\rm CDM}}(z).
\]
This residual form removes the dominant common \(\Lambda\)CDM background trend and makes the small model-dependent deviations easier to display. It also partially isolates the shape of the distance-redshift relation from the absolute supernova normalization.

For the fiducial flat \(\Lambda\)CDM background we use,

\begin{equation}
 E_{\Lambda{\rm CDM}}^2(z) = \Omega_{m0}(1 + z)^3 + \Omega_{\Lambda0}.
 \label{eq:E_LCDM}
\end{equation}
The supernova comparison is then displayed in terms of the residual distance modulus $\Delta \mu$ given above. In the diagnostic plot we remove a best-fit vertical offset from the Pantheon+ residuals relative to the same fiducial \(\Lambda\)CDM background.

\begin{figure}[htbp]
 \centering
 \includegraphics[width=0.82\textwidth]{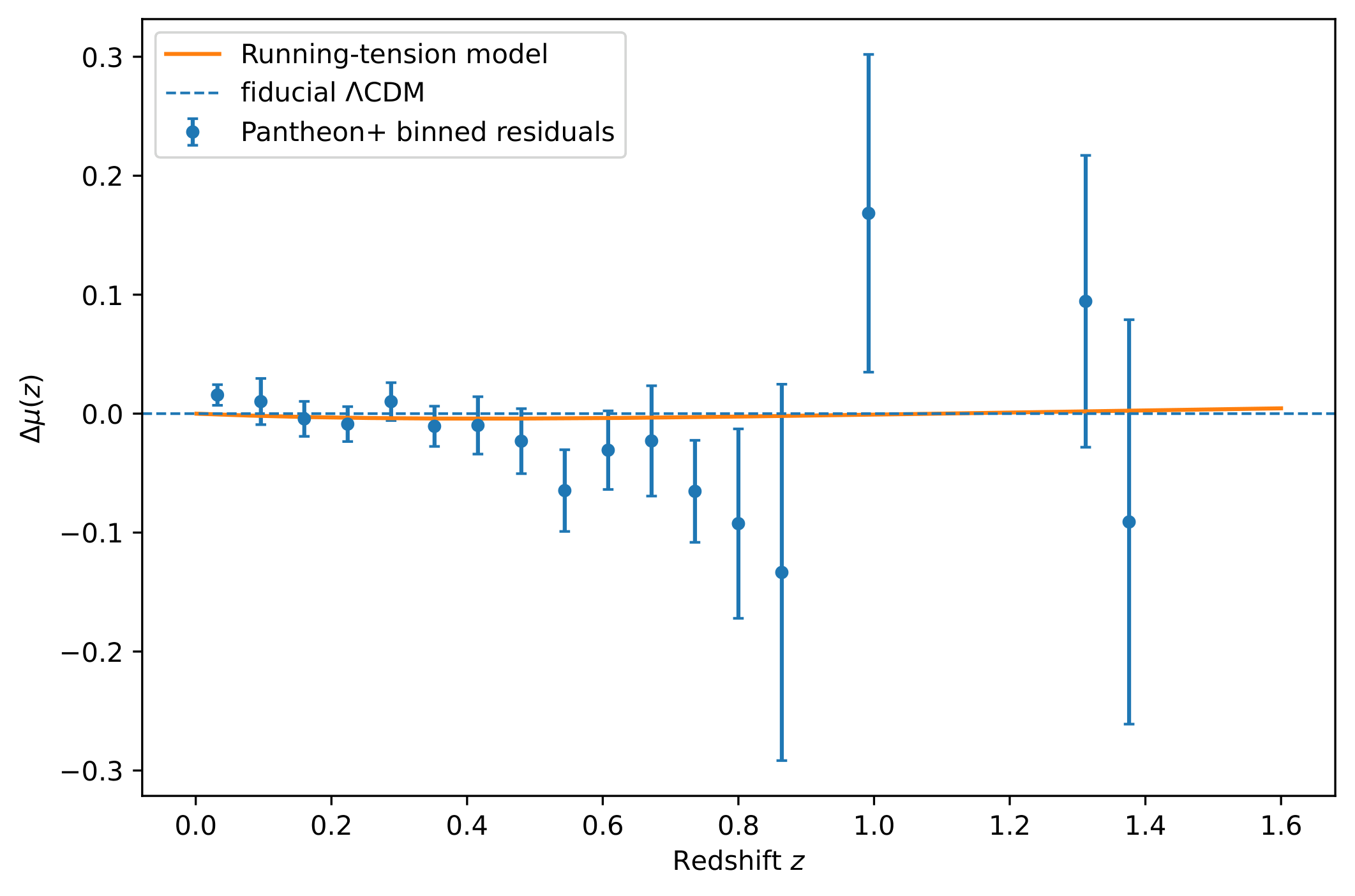}
 \caption{\footnotesize{Pantheon+ distance-modulus residual diagnostic. The points show binned Pantheon+ supernova residuals relative to a fiducial flat \(\Lambda\)CDM background, while the solid curve shows the prediction of the best-fit flux-tube running-tension model. The comparison is performed at the level of luminosity distance, not as a direct reconstruction of \(w(z)\).}}
 \label{fig:pantheon_residual_running_tension}
\end{figure}

Figure~\ref{fig:pantheon_residual_running_tension} shows the result. The Pantheon+ points are binned for visual clarity and are plotted as residuals with respect to fiducial \(\Lambda\)CDM. The running-tension curve remains close to the fiducial \(\Lambda\)CDM distance scale over the plotted range. This is expected because even a noticeable change in \(w(z)\) produces only an integrated and comparatively small effect in the luminosity distance. The figure should therefore be interpreted as a background-level consistency diagnostic, not as a full Pantheon+ likelihood analysis. In particular, the complete covariance matrix and all nuisance parameters have not been used to refit the model in this work.

\FloatBarrier
\noindent The main conclusion from Fig.~\ref{fig:pantheon_residual_running_tension} is that the running-tension model produces a supernova distance-modulus residual that is small compared with the scale of the current Pantheon+ residual scatter. The model therefore does not generate a large distortion of the luminosity-distance relation, despite producing a nontrivial low-redshift evolution of \(w_{\rm eff}^{\rm model}(z)\). This is important because it shows that the mechanism can generate evolving dark energy at the equation-of-state level while remaining close to the observed supernova distance-redshift relation at the background level.

At the same time, this plot should not be read as evidence of a full Pantheon+ fit. It is a diagnostic comparison in which the data are binned for visual clarity and a residual offset relative to fiducial \(\Lambda\)CDM is removed. A complete treatment would require the full Pantheon+ covariance matrix, nuisance-parameter marginalization, and a joint likelihood analysis with CMB and BAO constraints. Within the limited purpose of the present work, however, the Pantheon+ residual comparison supports the viability of the running-tension model as a proof-of-concept background cosmology.

\subsection{\centering Evolution of the Effective Dark-Energy Equation of State}
\label{subsec:weff_evolution}

\begin{figure}[htbp]
 \centering
 \includegraphics[width=0.82\textwidth]{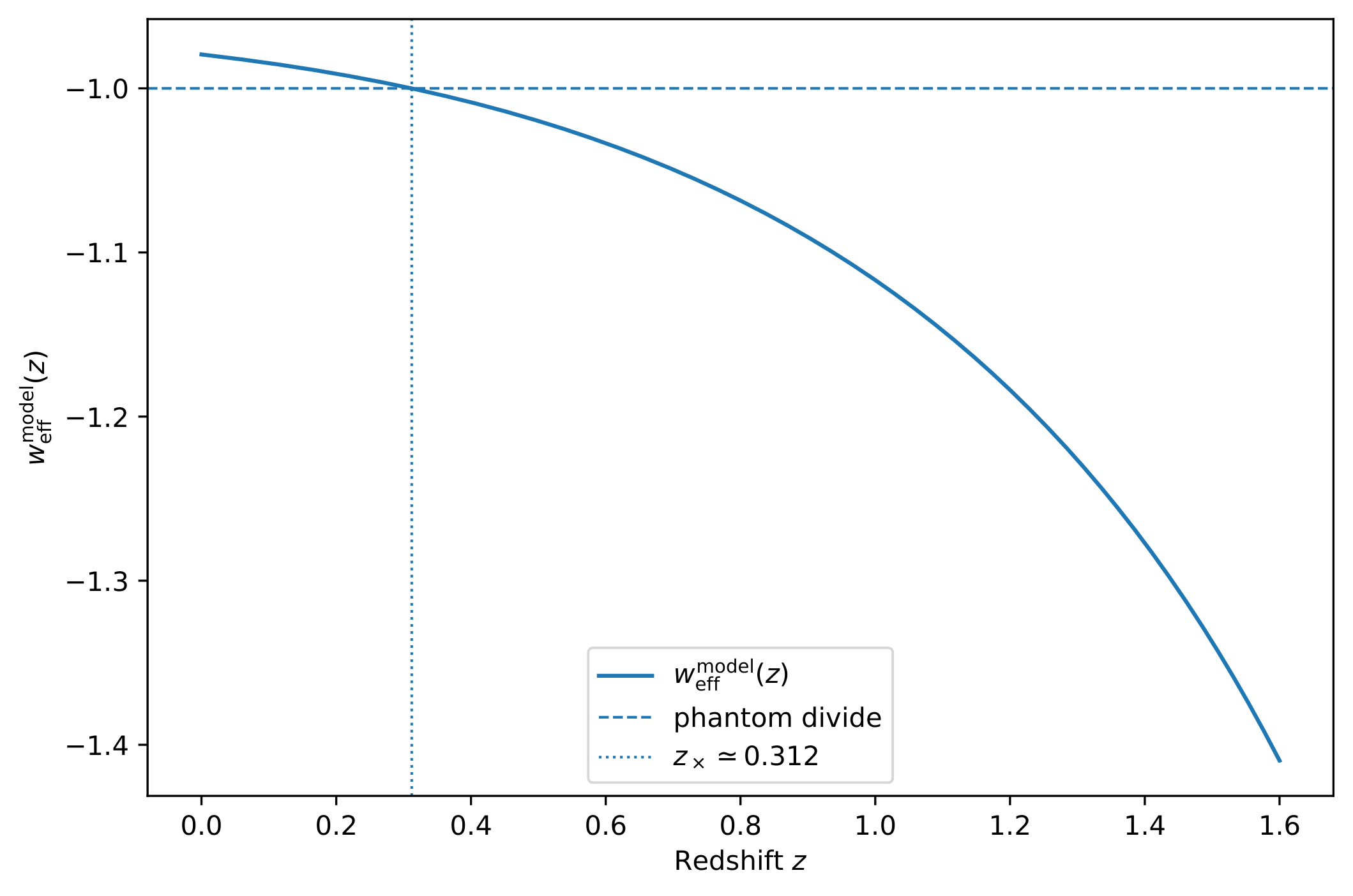}
 \caption{\footnotesize{Evolution of the effective dark-energy equation-of-state parameter in the best-fit running-tension model. The dashed horizontal line denotes the phantom divide \(w = - 1\), while the vertical dotted line marks the crossing redshift \(z_\times \simeq 0.312\). The crossing occurs when \(\eta(a_\times) = \kappa\), showing that it is generated dynamically by the hidden flux--tension exchange.}}
 \label{fig:weff_model_evolution}
\end{figure}

The previous comparisons show how the running-tension model behaves relative to compressed observational benchmarks. It is also useful to display the intrinsic evolution of the model equation of state itself. Figure~\ref{fig:weff_model_evolution} shows \(w_{\rm eff}^{\rm model}(z)\) for the final refined best-fit point,

\begin{equation}
 n = 2, \qquad a_i = \frac{1}{2.6}, \qquad \eta_i = 1, \qquad \Gamma = 1.45, \qquad \kappa = 0.1525.
\end{equation}
The model begins in the phantom regime shortly after the hidden transition, with,

\begin{equation}
 w_{\rm eff}^{\rm model}(z = 1.6) \simeq -1.4096,
\end{equation}
and evolves upward toward,

\begin{equation}
 w_{\rm eff}^{\rm model}(z = 0) \simeq -0.9793
\end{equation}
at the present epoch.

The phantom-divide crossing occurs when the hidden flux fraction reaches the balance value,

\begin{equation}
 \eta(a_\times) = \kappa.
\end{equation}
For the best-fit parameters this gives,

\begin{equation}
 z_\times \simeq 0.312.
\end{equation}
Thus the phantom crossing is not inserted by hand. It follows from the change in sign of the exchange term,

\begin{equation}
 Q = \Gamma H(\rho_{\rm flux} - \kappa T_s).
\end{equation}
At early times after the hidden transition, \(\eta>\kappa\), so energy flows from the flux-tube reservoir into the dynamical part of the tension sector and \(w_{\rm eff}^{\rm model}<-1\). At later times, the reservoir redshifts below the balance value, \(\eta<\kappa\), and the effective equation of state moves to \(w_{\rm eff}^{\rm model}>-1\).

\FloatBarrier

\subsection{\centering Comparison with Constraints on Different Dark-Energy Models}
\label{subsec:constraints_different_models}

We now place the running-tension result alongside the standard phenomenological descriptions of late-time acceleration. In \(\Lambda\)CDM, the dark-energy sector is exactly rigid and has,

\begin{equation}
 w = - 1.
\end{equation}
In the CPL plane this corresponds to the fixed point,

\begin{equation}
 (w_0, w_a) = (- 1, 0).
\end{equation}
In \(w\)CDM, the equation of state is allowed to differ from \(-1\), but it is still assumed to be constant,

\begin{equation}
 w(a) = w_0, \qquad w_a = 0 .
\end{equation}
Thus \(w\)CDM corresponds to the horizontal line \(w_a = 0\) in the CPL plane. In the full CPL or \(w_0w_a\)CDM parametrization, one allows the two-parameter form,

\begin{equation}
 w(a) = w_0 + w_a(1 - a) .
\end{equation}
The present running-tension model differs from these parametrized descriptions. Its equation of state is not assumed to be CPL from the outset. Instead, it is derived from the evolution of the total effective space tension,

\begin{equation}
 w_{\rm eff}^{\rm model}(a) = - 1 - \frac{1}{3} \, \frac{d\ln T(a)}{d\ln a}, \qquad T(a) = T_s + \delta T_s(a).
\end{equation}
Only after this physical curve is computed do we project it onto the CPL basis. Therefore the pair,

\begin{equation}
 (w_0^{\rm mod}, w_a^{\rm mod})
\end{equation}
should be interpreted as an effective CPL summary of the model curve, not as a fundamental parameter pair of the theory.

For the best-fit running-tension point found above, the effective CPL projection is,

\begin{equation}
 (w_0^{\rm mod}, w_a^{\rm mod}) = (-0.9153, -0.4572).
 \label{eq:running_tension_cpl_projection_final}
\end{equation}
This may be compared with the compressed DESI-CPL benchmark,

\begin{equation}
 (\hat w_0, \hat w_a) = (-0.828209, -0.744750).
 \label{eq:desi_cpl_mean_final}
\end{equation}
The difference between these two points is real, but the effective model projection lies in the same broad region of the CPL plane as the DESI-motivated benchmark.

To quantify this statement within the Gaussian approximation used in Sec.~4.1, define the squared distance,

\begin{equation}
 D^2(\theta) = (\theta - \hat{\theta})^T C^{-1}(\theta - \hat{\theta}), \qquad \theta =
 \begin{pmatrix}
  w_0 \\
  w_a
 \end{pmatrix}.
 \label{eq:gaussian_distance_cpl}
\end{equation}
For the running-tension projection one finds,

\begin{equation}
 D^2_{\rm mod} \simeq 2.09.
\end{equation}
For a two-dimensional Gaussian, the usual \(68.3\%\) contour corresponds to \(\Delta\chi^2 \simeq 2.30\), while the \(95.4\%\) contour corresponds to \(\Delta\chi^2 \simeq 6.17\). Thus, within the compressed Gaussian diagnostic used here, the running-tension projection lies close to the \(68\%\) contour of the DESI-CPL benchmark.

For comparison, the \(\Lambda\)CDM point gives,

\begin{equation}
 D^2_{\Lambda{\rm CDM}} \simeq 7.31
\end{equation}
relative to the same compressed benchmark. This does not mean that \(\Lambda\)CDM is ruled out by the present analysis, since the full posterior, dataset combinations, nuisance parameters, and model comparison machinery are not being reproduced here. It simply reflects the fact that the compressed DESI-motivated CPL benchmark used in this work is displaced from the exact \(\Lambda\)CDM point.

\begin{figure}[htbp]
 \centering
 \includegraphics[width=0.78\textwidth]{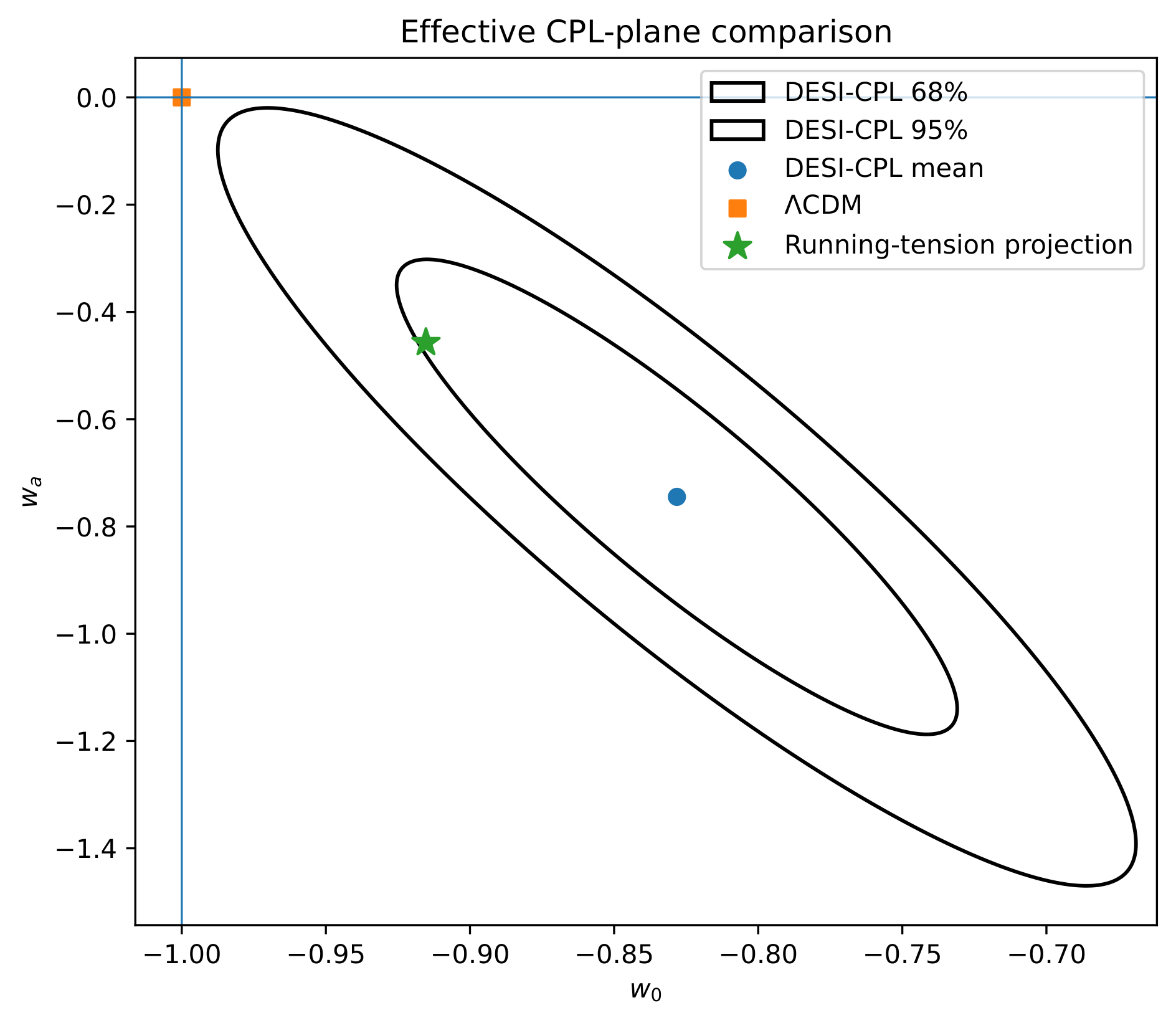}
 \caption{\footnotesize{Effective CPL-plane comparison. The blue point denotes the compressed DESI-CPL benchmark mean, while the ellipses show the corresponding Gaussian \(68\%\) and \(95\%\) contours obtained from the covariance matrix \(C\). The square denotes the \(\Lambda\)CDM point \((-1,0)\), and the star denotes the effective CPL projection of the best-fit running-tension model, \((w_0^{\rm mod},w_a^{\rm mod})=(-0.9153,-0.4572)\). The running-tension point should be interpreted as a projection of the model curve, not as a fundamental model parameter pair.}}
 \label{fig:w0wa_constraint_comparison}
\end{figure}

Figure~\ref{fig:w0wa_constraint_comparison} shows this comparison in the CPL plane. The blue point is the compressed DESI-CPL mean, the ellipses show the corresponding Gaussian \(68\%\) and \(95\%\) contours, the square marks \(\Lambda\)CDM, and the star marks the effective CPL projection of the running-tension model. The figure makes clear that the present model does not exactly reproduce the DESI-CPL mean, but it moves in the correct qualitative direction away from rigid \(\Lambda\)CDM and into the evolving dark-energy region preferred by the compressed benchmark.

\FloatBarrier
\noindent The comparison between different models is summarized in Table~\ref{tab:model_constraints_comparison}. The key distinction is that \(\Lambda\)CDM, \(w\)CDM, and CPL are parametrized descriptions of the dark-energy equation of state, whereas the present model derives \(w_{\rm eff}^{\rm model}(z)\) from the running of the space tension. The CPL pair quoted for the present model is therefore only a diagnostic projection.

\begin{table}[htbp]
 \centering
 \small
 \renewcommand{\arraystretch}{1.35}
 \setlength{\tabcolsep}{4pt}
 \begin{tabularx}{0.98\textwidth}{
   >{\raggedright\arraybackslash}p{0.16\textwidth}
   >{\raggedright\arraybackslash}X
   >{\raggedright\arraybackslash}X
   >{\raggedright\arraybackslash}X
  }
  \toprule
  Model 
  & Equation-of-state form 
  & Representative result in this comparison 
  & Interpretation \\
  \midrule
  
  \(\Lambda\)CDM
  &
  \(w = - 1\), or \((w_0, w_a) = (-1, 0)\)
  &
  \(D^2_{\Lambda{\rm CDM}} \simeq 7.31\)
  &
  Rigid cosmological constant
  \\[0.12cm]
  
  \(w\)CDM
  &
  \(w(a) = w_0\), with \(w_a = 0\)
  &
  One-dimensional subspace \(w_a = 0\)
  &
  Constant but non-\(\Lambda\) dark energy
  \\[0.12cm]
  
  CPL/\(w_0w_a\)CDM
  &
  \(w(a) = w_0 + w_a(1 - a)\)
  &
  \((\hat w_0, \hat w_a)\)\newline\(=(-0.828209, -0.744750)\)
  &
  DESI DR1 BAO + Pantheon+ + Planck 2018 compressed benchmark
  \\[0.12cm]
  
  Present model
  &
  \(\displaystyle w_{\rm eff} = - 1 - \frac{1}{3} \, \frac{d\ln T}{d\ln a}\)
  &
  \((w_0^{\rm mod}, w_a^{\rm mod}) = (-0.9153, -0.4572)\);
  \(D^2_{\rm mod} \simeq 2.09\)
  &
  CPL projection of the running-tension curve
  \\
  \bottomrule
 \end{tabularx}
 \caption{\footnotesize{Comparison between standard dark-energy parametrizations and the present running-tension model. The \(\Lambda\)CDM, \(w\)CDM, and CPL entries describe phenomenological equation-of-state parametrizations. The final row gives the effective CPL projection of the present model and should not be interpreted as a direct observational constraint on fundamental model parameters. The quantities \(D^2_{\Lambda{\rm CDM}}\) and \(D^2_{\rm mod}\) are computed using the compressed Gaussian benchmark of Sec.~4.1.}}
 \label{tab:model_constraints_comparison}
\end{table}
\FloatBarrier
\noindent The main result of this comparison is that the running-tension model moves away from the rigid \(\Lambda\)CDM point in the same qualitative direction as the compressed DESI-CPL benchmark. This comparison shows that the present model captures the qualitative direction of the DESI-motivated preference for evolving dark energy better than a strictly rigid \(\Lambda\)CDM equation of state within this simplified diagnostic.

\section{\centering Discussion, Limitations, and Future Work}
\label{sec:discussion_limitations_future}

The construction developed in this work should be understood as a proof of concept for a specific physical idea: the late-time dark-energy sector may be interpreted as the residual intrinsic tension of space, while mild departures from exact \(\Lambda\)CDM behavior may arise from a hidden defect reservoir that exchanges energy with the dynamical part of that tension. In this interpretation, the observed cosmological term is not replaced by an arbitrary dark-energy fluid. Instead, the uniform component of the space tension plays the role of the residual vacuum sector, while the running part \(\delta T_s(a)\) captures a late-time correction induced by hidden-sector dynamics.

The main dynamical result is that once the total tension is written as,

\begin{equation}
 T(a) = T_s + \delta T_s(a),
\end{equation}
the effective equation of state becomes,

\begin{equation}
 w_{\rm eff}(a) = - 1 - \frac{1}{3} \, \frac{d\ln T(a)}{d\ln a}.
\end{equation}
Thus any late-time exchange that changes \(\delta T_s(a)\) automatically generates a departure from \(w=-1\), even though the rigid background tension \(T_s\) remains untouched. In the present model this exchange is supplied by a hidden flux-tube reservoir produced after a late-time hidden transition \(U(1)_h\to\mathbb{Z}_n\). At the coarse-grained level, the exchange term,

\begin{equation}
 Q = \Gamma H(\rho_{\rm flux} - \kappa T_s)
\end{equation}
allows the model to cross the phantom divide dynamically. The crossing occurs when,

\begin{equation}
 \eta(a_\times) = \kappa,
\end{equation}
so that the sign of the energy transfer changes. This mechanism makes the phantom crossing a consequence of the flux--tension exchange, rather than an imposed feature of the equation of state.

The numerical comparison in Sec.~4 gives a useful phenomenological guide. Within the restricted scan considered here, the preferred representative point is,

\begin{equation}
 n = 2, \qquad a_i = \frac{1}{2.6}, \qquad \eta_i = 1, \qquad \Gamma = 1.45, \qquad \kappa = 0.1525.
\end{equation}
This corresponds to a string-like hidden reservoir with,

\begin{equation}
 w_{\rm flux} = \frac{n}{3} - 1 = - \frac{1}{3}.
\end{equation}
The resulting model curve evolves from,

\begin{equation}
 w_{\rm eff}^{\rm model}(z = 1.6) \simeq - 1.4096
\end{equation}
toward

\begin{equation}
 w_{\rm eff}^{\rm model}(z = 0) \simeq - 0.9793,
\end{equation}
with a phantom-divide crossing at,

\begin{equation}
 z_\times \simeq 0.312.
\end{equation}
When projected onto the CPL basis over \(0 \leq z \leq1.6\), the same curve gives,

\begin{equation}
 (w_0^{\rm mod}, w_a^{\rm mod}) \simeq (-0.9153, -0.4572).
\end{equation}
This does not exactly reproduce the compressed DESI-CPL benchmark,

\begin{equation}
 (\hat w_0, \hat w_a) = (-0.828209, -0.744750),
\end{equation}
but it lies in the same broad evolving-dark-energy region and produces a qualitatively similar low-redshift departure from \(\Lambda\)CDM.

The Pantheon+ comparison should be interpreted differently from the DESI-CPL comparison. The DESI-CPL benchmark is an effective equation-of-state comparison, whereas Pantheon+ constrains the luminosity-distance relation. For this reason, in Sec.~4.3 we translated the running-tension model into a background expansion function,

\begin{equation}
 E_{\rm mod}^2(z) = \Omega_{m0}(1 + z)^3 + \Omega_{{\rm DE}, 0} \frac{1 + \Delta(z)}{1 + \Delta(0)} ,
\end{equation}
and then into a luminosity distance and distance modulus. The residual comparison,

\begin{equation}
 \Delta\mu_{\rm mod}(z) = \mu_{\rm mod}(z) - \mu_{\Lambda{\rm CDM}}(z)
\end{equation}
shows that the best-fit running-tension curve remains close to the fiducial \(\Lambda\)CDM distance scale over the plotted redshift interval. This is expected because supernova distances are integrated quantities: even a visible evolution in \(w(z)\) can produce a relatively small effect in \(\mu(z)\). Therefore, the Pantheon+ plot should be viewed as a background-level consistency diagnostic. It indicates that the running-tension model is not obviously in conflict with the distance-modulus residual scale, but it does not constitute a full Pantheon+ likelihood analysis.

There are several limitations to the present treatment. First, the model does not solve the cosmological constant problem. The small residual baseline tension \(T_s\) is taken as an infrared input, assumed to survive after whatever ultraviolet mechanism neutralizes the large microscopic vacuum contributions. The present work addresses only whether that residual sector can acquire a mild late-time running.

Second, the exchange law has been treated phenomenologically. The coupling \(\Gamma\) is taken to be constant, and the flux-tube reservoir is described by a single coarse-grained energy fraction \(\eta(a)\) with a fixed effective dilution index \(n\). This is the simplest possible implementation, but a real hidden flux-tube network would generally have a richer evolution. Its energy density could depend on Hubble stretching, intercommutation, loop formation, loop decay, frictional damping, reconnection probability, hidden radiation losses, and possible interactions with the tension-sector scalar. These effects would generally make \(w_{\rm flux}\), \(\Gamma\), and even the effective balance parameter \(\kappa\) time dependent.

Third, the hidden transition has been modeled as an effective late-time onset rather than derived from a microscopic hidden-sector potential. In a more complete construction, the condensation of the hidden Higgs field should follow from an effective potential depending on temperature, curvature, the tension-sector scalar \(\sigma\), or other hidden-sector thresholds. Such a treatment would replace the abrupt transition at \(a_i\) by a finite-width transition and would determine the initial flux fraction \(\eta_i\) dynamically.

Fourth, the observational comparison is intentionally compressed. The DESI comparison uses a Gaussian approximation to the posterior in the \((w_0, w_a)\) plane, and the Pantheon+ plot is a residual diagnostic rather than a full covariance-level likelihood. A definitive test would require implementing the model in a Boltzmann solver and confronting it with the full combination of CMB, BAO, supernova, and large-scale-structure likelihoods.

These limitations also explain why a more refined calculation could plausibly improve the agreement with the observational benchmark. In the present minimal model, the shape of \(w_{\rm eff}(z)\) is controlled by only a few constants: \(n\), \(a_i\), \(\eta_i\), \(\Gamma\), and \(\kappa\). A realistic string-network treatment would introduce additional physical structure, not arbitrary curve fitting. For example, a distribution of string lengths, loop production rates, reconnection probabilities, and hidden radiation channels could modify the depletion history of \(\eta(a)\). A finite-width hidden transition could smooth the onset of the running tension. A coupling \(\Gamma(\sigma,a)\), derived from the same field \(\sigma\) that controls \(Z(\sigma)\) and \(v_\Phi(\sigma)\), could alter when and how efficiently energy is transferred between the hidden defect reservoir and the dynamical tension sector. These refinements could change the curvature of \(w_{\rm eff}(z)\), shift the phantom-crossing redshift, and move the effective CPL projection \((w_0^{\rm mod}, w_a^{\rm mod})\) closer to the observationally preferred region.

The most important future direction is therefore to replace the single-reservoir ansatz by a dynamical hidden string-network calculation. Such a calculation should track the formation of flux tubes after the \(U(1)_h \to \mathbb{Z}_n\) transition, the evolution of long strings and loops, the transfer of energy into hidden radiation or massive hidden modes, and the induced backreaction on \(\delta T_s(a)\). A second direction is to derive the exchange term \(Q\) directly from the effective action, rather than imposing it phenomenologically. A third direction is to study perturbations. Since the background equation of state can cross \(w = - 1\), one must check the stability of scalar perturbations, the effective sound speed, and possible signatures in growth-of-structure observables. Finally, a full likelihood-level comparison with CMB, BAO, Pantheon+, and structure-growth data will be necessary before the model can be regarded as more than a physically motivated proof of concept.

The central lesson of the present analysis is nevertheless clear. A residual vacuum-like tension of space can remain mostly rigid, while a hidden late-time defect reservoir induces a small dynamical correction. This provides a concrete and physically transparent route to mild running dark energy without requiring the baseline vacuum tension itself to decay.

\section{\centering Conclusions}
\label{sec:conclusions}

We have developed a phenomenological running-tension model of late-time dark energy. The starting point is the observation that an intrinsic uniform tension of space contributes to the gravitational field equations with the same tensor structure as vacuum energy. A Dirac--Born--Infeld completion of this intrinsic membrane picture naturally introduces a hidden Abelian gauge sector. If this hidden sector undergoes a late-time transition \(U(1)_h \to \mathbb{Z}_n\), hidden magnetic energy can be reorganized into a coarse-grained flux-tube reservoir.

The main result is that energy exchange between this hidden reservoir and the dynamical part of the space tension produces a time-dependent dark-energy density. The corresponding effective equation of state is,

\begin{equation}
 w_{\rm eff}(a) = - 1 - \frac{1}{3} \, \frac{d\ln T(a)}{d\ln a}, \qquad T(a) = T_s + \delta T_s(a).
\end{equation}
The rigid baseline \(T_s\) remains intact and only the correction \(\delta T_s(a)\) evolves. In this way the model realizes running dark energy without requiring the residual vacuum baseline itself to be removed.

A restricted numerical scan against a compressed DESI-CPL benchmark gives a representative best-fit point,

\begin{equation}
 n = 2, \qquad a_i = \frac{1}{2.6}, \qquad \eta_i = 1, \qquad \Gamma = 1.45, \qquad \kappa = 0.1525.
\end{equation}
This corresponds to a string-like hidden reservoir with \(w_{\rm flux} = - 1/3\). The resulting equation of state evolves from,

\begin{equation}
 w_{\rm eff}^{\rm model}(z = 1.6) \simeq -1.4096
\end{equation}
to
\begin{equation}
 w_{\rm eff}^{\rm model}(z = 0) \simeq -0.9793,
\end{equation}
and crosses the phantom divide at
\begin{equation}
 z_\times \simeq 0.312.
\end{equation}
The effective CPL projection of the model curve is
\begin{equation}
 (w_0^{\rm mod}, w_a^{\rm mod}) \simeq (-0.9153, -0.4572).
\end{equation}
This does not exactly reproduce the compressed DESI-CPL benchmark, but it demonstrates that a hidden flux-tube reservoir can generate a phenomenologically relevant low-redshift evolution of the dark-energy equation of state.

We also compared the same background evolution with Pantheon+ supernova distance information at the level of distance-modulus residuals. Since Pantheon+ constrains luminosity distances rather than directly measuring \(w(z)\), this comparison was treated as a diagnostic rather than a full likelihood analysis. The running-tension curve remains close to the fiducial \(\Lambda\)CDM distance scale over the plotted range, showing that the model is not obviously in conflict with the supernova residual scale at the background level.

The model should be regarded as a minimal proof of concept. It does not solve the cosmological constant problem, does not yet derive the hidden transition from first principles, and does not include a full dynamical simulation of the hidden string network or a full cosmological likelihood analysis. Nevertheless, it shows that once the residual vacuum sector is interpreted as an intrinsic tension of space, hidden-sector defect physics provides a natural mechanism for inducing mild running in the effective dark-energy component.

A more complete version of the framework should derive the flux--tension exchange from the hidden-sector action, model the evolution of the flux-tube network beyond the single-reservoir approximation, study perturbative stability and structure-growth signatures, and confront the model with full CMB, BAO, supernova, and large-scale-structure likelihoods. The present work therefore provides a starting point for a broader program in which late-time dark energy is treated not as an independent fluid, but as a dynamical manifestation of the intrinsic tension of space.

\section*{\centering Acknowledgment}
The author acknowledges the use of ChatGPT to assist with language editing and clarity in selected sections of this manuscript. The tool was used only to polish and rephrase text supplied by the author. All AI-assisted passages were reviewed and edited by the author, who exercised final editorial judgment. 

\section*{\centering Data and Code Availability}

The DESI-CPL benchmark used in this work was obtained from the publicly released DESI DR1 BAO cosmology posterior chains for the DESI DR1 BAO + Pantheon+ + Planck 2018 dataset combination. The Pantheon+ diagnostic uses the publicly available Pantheon+ supernova distance data. The custom Python scripts used to extract the Gaussian CPL benchmark, perform the restricted and refined flux-tube running-tension scans, compute the effective CPL projection, and generate the plots in Sec.~4 are available at Ref.~\cite{KhanMGKSPDataset_2}.

 \newpage
 
 \appendix
 
 \section{\centering Derivation of the effective equation of state}
 
 We now derive the relation,
 
 \begin{equation}
  w_{\rm eff}(a) =  - 1 - \frac{1}{3} \frac{d\ln T}{d\ln a} \, .
  \label{eq:weff_derivation_target}
 \end{equation}
 The starting point is the continuity equation for a homogeneous effective fluid in an FLRW background,
 
 \begin{equation}
  \dot{\rho}_{\rm DE} + 3H\bigl(\rho_{\rm DE} + p_{\rm DE}\bigr) = 0 \, ,
  \label{eq:continuity_DE}
 \end{equation}
 where $H = \dot a/a$ is the Hubble parameter. Writing the pressure in the usual form,
 
 \begin{equation}
  p_{\rm DE} = w_{\rm eff} \, \rho_{\rm DE} \, ,
  \label{eq:pweff}
 \end{equation}
 one finds,
 
 \begin{equation}
  \dot{\rho}_{\rm DE} + 3H\bigl(1 + w_{\rm eff}\bigr) \, \rho_{\rm DE} = 0 \, .
  \label{eq:continuity_weff}
 \end{equation}
 Solving for $w_{\rm eff}$ gives,
 
 \begin{equation}
  w_{\rm eff} =  - 1 - \frac{\dot{\rho}_{\rm DE}}{3H\rho_{\rm DE}} \, .
  \label{eq:weff_rho_dot}
 \end{equation}
 Using,
 
 \begin{equation}
  \frac{d\ln \rho_{\rm DE}}{d\ln a} = \frac{1}{\rho_{\rm DE}} \, \frac{d\rho_{\rm DE}}{d\ln a} = \frac{\dot{\rho}_{\rm DE}}{H\rho_{\rm DE}} \, ,
  \label{eq:dlnrho_dloga}
 \end{equation}
 Equation \eqref{eq:weff_rho_dot} therefore becomes,
 
 \begin{equation}
  w_{\rm eff}(a) = - 1 - \frac{1}{3} \frac{d\ln \rho_{\rm DE}}{d\ln a}.
  \label{eq:weff_rho_general}
 \end{equation}
 In the present framework, the effective dark-energy density is identified with the tension of space,
 
 \begin{equation}
  \rho_{\rm DE}(a) \equiv T(a)\, .
  \label{eq:rhoDE_equals_Ts}
 \end{equation}
 Therefore,
 
 \begin{equation}
  w_{\rm eff}(a) = - 1 - \frac{1}{3} \frac{d\ln T}{d\ln a} \, .
  \label{eq:weff_Ts_final}
 \end{equation}
 If we further split the tension as,
 
 \begin{equation}
  T(a) = T_s + \delta T_s(a) \, , \qquad \dot T_s = 0 \, ,
  \label{eq:Ts_split_derivation}
 \end{equation}
 then,
 
 \begin{equation}
  \frac{d\ln T}{d\ln a} = \frac{1}{T_s + \delta T_s} \frac{d\delta T_s}{d\ln a} \, .
  \label{eq:dlnTs_split}
 \end{equation}
 Substituting this into Eq.~\eqref{eq:weff_Ts_final}, one obtains,
 
 \begin{equation}
  w_{\rm eff}(a) = - 1 - \frac{1}{3\bigl(T_s + \delta T_s\bigr)} \frac{d\delta T_s}{d\ln a} \, .
  \label{eq:weff_deltaTs_final}
 \end{equation}

\newpage

\section{Derivation of the coupled continuity equations}

We model the late-time interacting sector in coarse-grained form as the sum of two components, fist of which is a hidden flux-tube reservoir and second being the dynamical part of the space-tension sector. Accordingly, we split the total stress tensor as,

\begin{equation}
 T^{\mu\nu}_{\rm tot} = T^{\mu\nu}_{\rm flux} + T^{\mu\nu}_{\delta T_s} \, .
 \label{eq:Ttot_split}
\end{equation}
We assume that the total interacting sector is covariantly conserved,

\begin{equation}
 \nabla_\mu T^{\mu\nu}_{\rm tot} = 0 \, ,
 \label{eq:total_cov_cons}
\end{equation}
but that the two individual components may exchange energy through an
interaction current $Q^\nu$ according to,

\begin{equation}
 \nabla_\mu T^{\mu\nu}_{\rm flux} = - Q^\nu \, , \qquad \nabla_\mu T^{\mu\nu}_{\delta T_s} = + Q^\nu \, .
 \label{eq:split_cov_cons}
\end{equation}
For a homogeneous FLRW background, the transfer is taken to be purely temporal in the comoving frame,

\begin{equation}
 Q^\nu = Q \, u^\nu \, , 
 \label{eq:Qnu_def}
\end{equation}
where $u^\nu$ is the background four-velocity and $Q$ is the scalar energy-transfer rate.

The flux-tube sector is described as an effective fluid with,

\begin{equation}
 p_{\rm flux}=w_{\rm flux}\,\rho_{\rm flux} \, .
 \label{eq:pflux_wflux}
\end{equation}
Its continuity equation then becomes,

\begin{equation}
 \dot{\rho}_{\rm flux} + 3 H \bigl(\rho_{\rm flux} + p_{\rm flux}\bigr) =
 - \, Q \, ,
 \label{eq:flux_cont_pre}
\end{equation}
or equivalently,

\begin{equation}
 \dot{\rho}_{\rm flux} + 3 H \bigl(1 + w_{\rm flux}\bigr)\rho_{\rm flux} = - \, Q \, .
 \label{eq:rhoflux_cont_derived}
\end{equation}
We now treat the dynamical transferred part of the space tension as a vacuum-like component with,

\begin{equation}
 \rho_{\delta T_s} = \delta T_s \, , \qquad p_{\delta T_s} = - \, \delta T_s \, .
 \label{eq:deltaTs_vaclike}
\end{equation}
Its continuity equation is therefore,

\begin{equation}
 \dot{\rho}_{\delta T_s} + 3 H \bigl(\rho_{\delta T_s} + p_{\delta T_s} \bigr) = Q \, .
 \label{eq:deltaTs_cont_pre}
\end{equation}
Since,

\begin{equation}
 \rho_{\delta T_s} + p_{\delta T_s} = 0 \, ,
\end{equation}
this reduces immediately to,

\begin{equation}
 \dot{\delta T_s} = Q \, .
 \label{eq:deltaTs_cont_derived}
\end{equation}
We thus obtain the coupled background system,

\begin{align}
 \dot{\rho}_{\rm flux} + 3 H \bigl(1 + w_{\rm flux}\bigr)\rho_{\rm flux}
 & = - \, Q \, , 
 \\[1ex]
 \dot{\delta T_s} & = Q \, .
\end{align}
These equations simply express that the hidden flux-tube sector loses exactly the amount of energy gained by the dynamical part of the space-tension sector, and conversely when the sign of $Q$ reverses.

\newpage

\section{\centering Redshift-Space Form of the Running-Tension Equation of State}

In the main text, the effective dark-energy equation of state of the running-tension sector is written as,

\begin{equation}
 w_{\rm eff}^{\rm model}(a) = - 1 - \frac{1}{3} \, \frac{d\ln T(a)}{d\ln a}, \qquad T(a) = T_s + \delta T_s(a).
 \label{eq:app_weff_a_start}
\end{equation}
This appendix derives the corresponding expression as a function of redshift \(z\).

The relation between scale factor and redshift is,

\begin{equation}
 a = \frac{1}{1 + z}, \qquad \ln a = - \ln(1 + z).
 \label{eq:app_a_z_relation}
\end{equation}
Therefore,

\begin{equation}
 \frac{d\ln a}{dz} = - \frac{1}{1 + z},
\end{equation}
which implies,

\begin{equation}
 \frac{d}{d\ln a} = - (1 + z)\frac{d}{dz}.
 \label{eq:app_derivative_conversion}
\end{equation}
Using this in Eq.~\eqref{eq:app_weff_a_start}, one obtains,

\begin{equation}
 w_{\rm eff}^{\rm model}(z) = - 1 + \frac{1 + z}{3} \frac{d\ln T(z)}{dz}.
 \label{eq:app_weff_z_logT}
\end{equation}
Since,

\begin{equation}
 T(z) = T_s + \delta T_s(z),
\end{equation}
and \(T_s\) is constant, this gives,

\begin{equation}
 w_{\rm eff}^{\rm model}(z) = - 1 + \frac{1 + z}{3\left[T_s + \delta T_s(z)\right]} \frac{d\delta T_s(z)}{dz}.
 \label{eq:app_weff_z_deltaTs}
\end{equation}
It is useful to introduce the dimensionless running-tension variable,

\begin{equation}
 \Delta(z) \equiv \frac{\delta T_s(z)}{T_s}.
 \label{eq:app_Delta_def}
\end{equation}
Then,

\begin{equation}
 T(z) = T_s\left[1 + \Delta(z)\right],
\end{equation}
and Eq.~\eqref{eq:app_weff_z_deltaTs} becomes,

\begin{equation}
 w_{\rm eff}^{\rm model}(z) = - 1 + \frac{1 + z}{3\left[1 + \Delta(z)\right]} \frac{d\Delta(z)}{dz}.
 \label{eq:app_weff_z_general}
\end{equation}
This is the general redshift-space form of the running-tension equation of state.

For the flux-tube reservoir model considered in the main text, the dimensionless transferred tension satisfies,

\begin{equation}
 \frac{d\Delta}{d\ln a} = \Gamma\left[\eta(a) - \kappa\right],
 \label{eq:app_Delta_evolution_a}
\end{equation}
where \(\eta(a)\) is the hidden flux fraction, \(\Gamma\) is the dimensionless exchange strength, and \(\kappa\) is the balance parameter. The hidden flux fraction is,

\begin{equation}
 \eta(a) = \eta_\infty + \left(\eta_i - \eta_\infty\right) \left(\frac{a}{a_i}\right)^{-m},
 \label{eq:app_eta_a}
\end{equation}
with,

\begin{equation}
 m = n + \Gamma, \qquad n = 3(1 + w_{\rm flux}), \qquad \eta_\infty = \frac{\Gamma \kappa}{m}.
 \label{eq:app_m_eta_infty}
\end{equation}
Here \(a_i\) is the scale factor at which the hidden transition occurs, and \(\eta_i\) is the corresponding initial hidden flux fraction.

Using \(a=(1+z)^{-1}\), one has,

\begin{equation}
 \frac{a}{a_i} = \frac{1}{a_i(1 + z)}.
\end{equation}
Therefore,

\begin{equation}
 \left(\frac{a}{a_i}\right)^{-m} = \left[a_i(1 + z)\right]^m,
\end{equation}
and hence,

\begin{equation}
 \eta(z) = \eta_\infty + \left(\eta_i - \eta_\infty\right) \left[a_i(1 + z)\right]^m.
 \label{eq:app_eta_z}
\end{equation}

We now integrate Eq.~\eqref{eq:app_Delta_evolution_a}. Let,

\begin{equation}
 N \equiv \ln a, \qquad N_i \equiv \ln a_i.
\end{equation}
With the initial condition,

\begin{equation}
 \Delta(a_i)=0,
\end{equation}
one finds,

\begin{align}
 \Delta(a) & = \Gamma \int_{N_i}^{N} \left[\eta(N') - \kappa\right]dN'
 \nonumber\\
 & = \Gamma \int_{N_i}^{N} \left[ \eta_\infty - \kappa + \left(\eta_i - \eta_\infty\right) e^{-m(N'-N_i)} \right]dN'.
\end{align}
Performing the integral gives,

\begin{equation}
 \Delta(a) = \Gamma \left[ (\eta_\infty - \kappa) \ln \, \left(\frac{a}{a_i}\right) + \frac{\eta_i - \eta_\infty}{m} \left\{
 1 - \left(\frac{a}{a_i}\right)^{-m} \right\} \right].
 \label{eq:app_Delta_a}
\end{equation}
Converting this expression to redshift gives,

\begin{equation}
 \ln\left(\frac{a}{a_i}\right) = - \ln\left[a_i(1 + z)\right],
\end{equation}
and

\begin{equation}
 \left(\frac{a}{a_i}\right)^{-m} = \left[a_i(1 + z)\right]^m.
\end{equation}
Therefore,

\begin{equation}
 \Delta(z) = \Gamma \left[ (\kappa - \eta_\infty) \ln\left[a_i(1 + z)\right] + \frac{\eta_i - \eta_\infty}{m} \left\{ 1 - \left[a_i(1 + z)\right]^m \right\} \right].
 \label{eq:app_Delta_z}
\end{equation}

Differentiating Eq.~\eqref{eq:app_Delta_z} with respect to \(z\), we obtain,

\begin{align}
 \frac{d\Delta}{dz} & = \Gamma \left[ \frac{\kappa - \eta_\infty}{1 + z} - \left(\eta_i - \eta_\infty\right) a_i^m(1 + z)^{m - 1} \right]
 \nonumber\\
 & = \frac{\Gamma}{1 + z} \left[ \kappa - \eta_\infty - \left(\eta_i - \eta_\infty\right) \left[a_i(1 + z)\right]^m
 \right].
\end{align}
Using Eq.~\eqref{eq:app_eta_z}, this simplifies to,

\begin{equation}
 \frac{d\Delta}{dz} = - \frac{\Gamma}{1 + z} \left[\eta(z) - \kappa\right].
 \label{eq:app_dDelta_dz}
\end{equation}
Substituting Eq.~\eqref{eq:app_dDelta_dz} into Eq.~\eqref{eq:app_weff_z_general}, we obtain,

\begin{equation}
 w_{\rm eff}^{\rm model}(z) = - 1 - \frac{ \Gamma \left[\eta(z) - \kappa\right] }{ 3 \left[1 + \Delta(z)\right]}.
 \label{eq:app_weff_z_flux}
\end{equation}
This is the redshift-space equation of state used for the numerical comparison in the main text.

Substituting Eqs.~\eqref{eq:app_eta_z} and \eqref{eq:app_Delta_z} into Eq.~\eqref{eq:app_weff_z_flux}, the fully explicit expression is

\begin{equation}
\begin{aligned}
 w_{\rm eff}^{\rm model}(z) = -1 -
 \frac{\Gamma\left[\eta_\infty + (\eta_i-\eta_\infty)\left[a_i(1+z)\right]^m - \kappa\right]}
 {3\left[1 + \Gamma\left\{(\kappa-\eta_\infty)\ln\left[a_i(1+z)\right]
 + \dfrac{\eta_i-\eta_\infty}{m}\left(1-\left[a_i(1+z)\right]^m\right)\right\}\right]} \, .
\end{aligned}
 \label{eq:weff_model_z_explicit}
\end{equation}
The expression above applies after the hidden transition has occurred, namely for,

\begin{equation}
 0 \leq z \leq z_i, \qquad z_i = \frac{1}{a_i} - 1.
\end{equation}
For \(z>z_i\), if the hidden flux-tube reservoir has not yet formed, the model reduces to the rigid-tension limit,

\begin{equation}
 w_{\rm eff}^{\rm model}(z) = - 1 .
\end{equation}
Thus the departure from \(\Lambda\)CDM begins only after the late hidden-sector transition.

\end{document}